\documentclass[11pt,preprint,superscriptaddress,amsmath,amssymb,11pt,aps,showpacs,showkeys,nofootinbib,a4paper]{article}
\usepackage{amssymb,amsmath,amsfonts}
\usepackage{graphicx}
\usepackage{eepic,epsfig}
\usepackage{verbatim}
\usepackage{color}
\usepackage{hyperref}

1.60cm \makeatletter \@addtoreset{equation}{section}

\makeatother

\begin{document}
\title{Vacuum polarization in high-dimensional AdS spacetime in the presence of a cosmic string and a compactified extra dimension}
\author{W. Oliveira dos Santos\thanks {E-mail: wagner.physics@gmail.com}, E. R. Bezerra de Mello\thanks
{E-mail: emello@fisica.ufpb.br} and H. F. Mota\thanks{E-mail: hmota@fisica.ufpb.br}\\
\\
\textit{Departamento de F\'{\i}sica, Universidade Federal da Para\'{\i}ba}\\
\textit{58059-900, Caixa Postal 5008, Jo\~{a}o Pessoa, PB, Brazil.}\vspace{%
0.3cm}\\
}
\maketitle
%
\begin{abstract}
In this paper, we analyze the vacuum expectation values (VEVs) of the
field squared and the energy–momentum tensor associated with a charged and massive scalar quantum field in a generalized $(D+1)$-dimensional anti-de Sitter space in the presence of a cosmic string, admitting a magnetic flux running along the string's core. In addition we admit that an extra coordinate is compactified to a circle and presents an extra magnetic flux running along its center. This compactification is implemented by imposing a quasiperiodic condition on the field with an arbitrary phase. The calculation of the VEVs of the field squared and the energy-momentum tensor, are developed by using the positive-energy Wightman function. The latter is constructed  by the mode sum over the complete set of normalized bosonic wave-functions. Due to the compactification, the Wightman function is presented as the sum of two distinct contributions. The first one corresponding to the idealized cosmic string, and the second is induced by the compactification. The latter goes to zero for an infinite length of the compactification. As a consequence of the general structure of the Wightman function, both VEVs present also this decomposition. Moreover, due to the Aharanov-Bohm type of interaction  between the field with the magnetic fluxes, the VEVs depend on the fractional part of the ration between the total flux and the quantum one.
\end{abstract}
\bigskip
%
\section{Introduction}
\label{Int}
%
It is believed that the Universe has been underwent several symmetry breaking phase transitions during its expansion. In these transitions different types of topological objects may have been created, like global monopoles and cosmic strings \cite{Kibble,V-S}. In particular, cosmic strings are of special interest. Although recent observational data on the cosmic microwave background radiation have ruled out cosmic strings as the primary source for primordial density
perturbation, they are still candidate for the generation of a number of
interesting physical effects like gamma ray bursts \cite{Berezinski},
gravitational waves \cite{Damour} and high energy cosmic rays \cite{Bhattacharjee}. In addition cosmic string has also been considered in scenarios beyond the standard model of particle physics, like in supersymmetry and in string theory approaches \cite{Hind,CopelandJ}.  
Phenomenologically, current observations of CMB suggest cosmic strings can contribute to a small percentage of the primordial density perturbations \cite{Ade:2013xla} in the Universe.  More recently, cosmic strings have attracted renewed interest due to a variant of their formation mechanism that is proposed in the framework of brane inflation \cite{Sarangi}-\cite{Dvali}.

The geometry of the spacetime produced by a cosmic string can be approximately described by a planar angle deficit on the two-dimensional sub-space perpendicular to it \cite{Vile81,VS}. Although this object was first introduced in the literature as being created by a Dirac-delta type distribution of energy and axial stress  along a straight infinity line, it can also  be described by classical field theory where the energy-momentum tensor associated with the Maxwell-Higgs  system, investigated by Nielsen and Olesen in \cite{Nielsen197345}, couples to the Einstein's equations. This coupled system was first investigated in \cite{PhysRevD.32.1323} and \cite{Linet1987240}.

The conical structure associated with a cosmic string provides relevant effects in quantum field theories, inducing non-zero vacuum expectation values for physical observables. In fact the analysis of the vacuum
polarization effects associated to scalar and fermionic fields have been
developed in \cite{scalar}-\cite{Site11} and \cite{ferm}-\cite{Beze08f},
respectively, for the geometry of an idealized cosmic string on a flat background. On the other hand, in curved backgrounds additional effects appear due to the bulk gravitational field. Recently the combined effects of non-trivial
topology and background curvature on the local characteristics of the scalar and fermionic vacua have been investigated in \cite{StringdS,StringdS1} and \cite{StringAdS,StringAdS1} for a cosmic string in de Sitter and anti-de Sitter spacetimes, respectively. 

 Another type of topological quantum effect appears in models with compact spatial dimensions. The presence of compact dimensions is a key feature of 	most high-energy theories of fundamental physics. In the model introduced by Randall and Sundrum \cite{RS} the background
geometry consists of two parallel branes, with positive and negative tensions, embedded in a five-dimensional AdS bulk. The fifth coordinate is compactified on
orbifold and the branes are located at the two fixed points. Another interesting application of the field theoretical models with compact dimensions
appeared in nanophysics. The long-wavelength description of the electronic states in graphene can be formulated in terms of the Dirac-like theory in three-dimensional spacetime with the Fermi velocity playing the role of speed of light (see, e.g., \cite{Castro}). Single-walled carbon nanotubes are generated by rolling up a graphene sheet to form a cylinder and the background spacetime for the corresponding Dirac-like theory has the topology $R^2\times S^1$. Also the Casimir effects inspired in by nanotubes has been calculated for scalar field obeying quasiperiodical boundary condition in \cite{Feng}.

AdS spacetime is remarkable from different points of view \cite{Soko}. The first interest was motivated by the fact that the quantum analysis of fields  propagating on this curved space can be developed exactly. The
presence of both regular and irregular modes and the possibility of
interesting causal structure lead to a number of new phenomena. The
importance of this theoretical work increased when it was discovered that
AdS spacetime generically arises as a ground state in extended supergravity, in string theories and braneworld scenario \cite{Perez,Amaro}. 

The analysis of an induced current associated with a charged scalar quantum field in a high-dimensional AdS spacetime in the presence of a cosmic string  taking into account also the presence of a magnetic flux running along the string's core, was developed in \cite{Santos:2018ttf}. There, in order to analyze a more general situation, a compactification along an extra dimension was considered.\footnote{The presence of compactification of spatial dimensions serves to alter vacuum fluctuations of a quantum field and leads to the Casimir-type contributions in the vacuum expectation values of physical observables (see Refs. \cite{Most97,Eliz94} for the topological Casimir effect and its role in cosmology).}   Here in this paper we would like to continue along this line of investigation, and study the vacuum expectation value (VEV) of the energy-momentum tensor. This analysis will be developed adopting a similar procedure as exhibited in \cite{Santos:2018ttf}. 

The outline of this paper is the following: In the section \ref{sec2} we provide the geometry of the spacetime to be considered, and the obtainment of the complete set of normalized solutions of the Klein-Gordon equation, by using \textit{Poincar\'{e}} coordinates and admitting an arbitrary curvature coupling.
In addition we consider the presence of a azimuthal and axial vector potentials. The later introduced as a new ingredient due to the compactification. Having the set of wave-function we construct the positive-energy Wightaman function. In section  \ref{sec3}  we explicitly calculate the VEV of the field squared. We show that this VEV is decomposed in three parts, the first one induced by the AdS curved space, followed by the contribution induced by the cosmic string and the compactification. In section \ref{sec4} we calculate the VEV of the energy-momentum tensor. The later presents similar structure as the field squared. In Section \ref{concl} we summarize our most relevant results. In appendix \ref{Apendx} we present the derivation of a compact expression for the calculation of the energy-momentum tensor. In this paper we shall use the units $\hbar =G=c=1$.

\section{Klein-Gordon equation and Wightman function}
\label{sec2}
In this section, we briefly discuss the evaluation of the Wightman function found in \cite{Santos:2018ttf}. So, let us consider a charged scalar quantum field, $\varphi(x)$, on the background of a $(D+1)$-dimensional AdS spacetime, with $D>3$, in presence of a cosmic string and a compactified extra dimension. The analysis of the vacuum polarization effects associated with a scalar field is generally made by the use of the positive frequency Wightman function. In order to evaluate this function, we first obtain the complete set of normalized mode functions for the Klein-Gordon equation admitting an arbitrary curvature coupling parameter.

By using cylindrical coordinates system, the geometry associated with a cosmic string in a $(3+1)$-dimensional AdS spacetime is given by the line element below:
\begin{equation}
ds^{2}=e^{-2y/a}[dt^{2}-dr^{2}-r^{2}d\phi ^{2}]-dy^{2}\ ,  \label{ds1}
\end{equation}
where $r\geqslant 0$ and $\phi \in \lbrack 0,\ 2\pi /q]$ define the
coordinates on the conical geometry, $(t, \ y)\in (-\infty ,\ \infty )$, and
the parameter $a$ determines the curvature scale of the background
spacetime. In the above coordinate system the string is along the $y-$axis.  The parameter $q\geq 1$ defines the planar angle deficit on the two-dimensional surface orthogonal to the string. Using the \textit{Poincar\'{e}} coordinate defined by $w=ae^{y/a}$, the line element above can be conformally related to the line element associated with a cosmic string in Minkowski spacetime:
\begin{equation}
ds^2 = \left(\frac{a}{w}\right)^2[dt^2 - dr^2 - r^2d\varphi^2 - dw^2 ] \ .
\label{ds2}
\end{equation}
For the new coordinate one has $w\in \lbrack 0,\ \infty )$. Specific values for this coordinates deserve to be mentioned: $w=0$ and $w=\infty $ correspond to the AdS boundary and horizon, respectively.

To generalize the line element (\ref{ds2}) to $(D+1)$-dimensional, with $D>3$, AdS spacetimes, we adopt the standard procedure, by adding extra Euclidean coordinates \cite{deMello:2011ji}:
\begin{equation}
ds^2 = \left(\frac{a}{w}\right)^2\bigg[dt^2 - dr^2 - r^2d\varphi^2 - dw^2 - \sum_{i=4}^{D}(dx^i)^2\bigg]   \   .
\end{equation}
In fact, the Euclidian version of the line element inside the bracket has been proposed by Linet in \cite{Linet}.

The cosmological constant, $\Lambda $, and the Ricci scalar, $R$, are related with the scale $a$ by the formulas
\begin{equation}
\Lambda =-\frac{D(D-1)}{2a^{2}} \ ,\ \ R=-\frac{D(D+1)}{a^{2}}\ .
\label{LamR}
\end{equation}

Since we are interested in investigating the effect of a charged scalar field coupled to a gauge field on the vacuum polarization, we admit the presence of magnetic flux running along the string's core. Moreover, we also assume the compactification along only one extra coordinate, defined by $z$ in the expression below, 
\begin{equation}
ds^2 = \left(\frac{a}{w}\right)^2\bigg[dt^2 - dr^2 - r^2d\phi^2 - dw^2 -dz^2- \sum_{i=5}^{D}(dx^i)^2\bigg]   \   .
\label{le2}
\end{equation}
Note that we will also consider the presence of a constant vector potential along the extra compact dimension. This compactification is implemented by assuming that $z\in[0, \ L]$, and the matter field obeys the quasiperiodicity condition below,
\begin{equation}
\varphi(t,r,\phi, w, z + L, x^5,...,x^{D}) = e^{2\pi i\beta}\varphi(t,r,\phi, w, z, x^5,...,x^{D}),
\label{QPC}
\end{equation}
where $0\leq\beta\leq 1$.The special cases $\beta =0$ and $\beta =1/2$ correspond to the untwisted and twisted fields, respectively, along the $z$-direction. 

The field equation which governs the quantum dynamics of a charged
bosonic field with mass $m$, in a curved background and in the presence of an 
electromagnetic potential vector, $A_\mu$, reads
\begin{equation}
(g^{\mu\nu}D_{\mu}D_{\nu} + m^2 + \xi R)\varphi(x) = 0  \   , 
\label{KGE}
\end{equation}
being $D_{\mu}=\partial_{\mu}+ieA_{\mu}$. In addition, we have considered the presence of a non-minimal coupling, $\xi$, between the field and the geometry represented by the Ricci scalar, $R$. Two specific values for the curvature coupling are $\xi = 0$ and $\xi = \frac{D - 1}{4D}$, that correspond to minimal and conformal coupling, respectively. Also we shall assume the existence of the following constant vector potentials,
\begin{equation}
A_{\mu} = (0,0,A_{\phi}, 0, A_z, 0,...,0)  \  ,
\label{VP}
\end{equation}
with $A_{\phi}=-q\Phi_\phi/(2\pi)$ and $A_{z}=-\Phi_z/L$, being $\Phi_\phi$ and $\Phi_z$
the corresponding magnetic fluxes. In quantum field  theory the  condition
\eqref{QPC} changes the spectrum of the vacuum fluctuations compared with the case of uncompactified dimension and, as a consequence, the induced vacuum current density changes. 

In the spacetime defined by \eqref{le2} and in the presence of the vector 
potentials given above, the equation \eqref{KGE} becomes
\begin{eqnarray}
\left[\frac{\partial^2}{\partial t^2} - \frac{\partial^2}{\partial r^2} - \frac{1}{r}\frac{\partial}{\partial r} - \frac{1}{r^2}\left(\frac{\partial}{\partial\phi} + ieA_{\phi}\right)^2 -
\left(\frac{\partial}{\partial z} + ieA_{z}\right)^2\right.\nonumber\\
\left.
- \frac{\partial^2}{\partial w^2}-\frac{(1-D)}{w}\frac{\partial}{\partial w} + \frac{M(D,m,\xi)}{w^2} - \sum_{i=5}^{D}\frac{\partial^2}{\partial (x^i)^2} \right]\varphi(x) = 0  \  . 
\label{KGE2}
\end{eqnarray}
where $M(D,m,\xi) = a^2m^2 - \xi D(D+1)$. 

The equation above is completely separable and its positive energy and regular solution at origin is given by,
\begin{equation}
\varphi(x) = Cw^{\frac{D}{2}}J_{\nu}(pw)J_{q|n +\alpha|}(\lambda r)e^{-iE t + iqn\phi + ik_{z}z + i\vec{k}\cdot\vec{x}_{\parallel}}.
\label{Solu1}
\end{equation}
In the expression above $\vec{x}_{\parallel}$ represents the coordinates along the $(D-4)$ extra dimensions, and $\vec{k}$ the corresponding momentum. Moreover,
\begin{eqnarray}
\nu &=& \sqrt{\frac{D^2}{4} + a^2m^2 - \xi D(D+1)},\nonumber\\
E &=& \sqrt{\lambda^2 + p^2 + \vec{k}^2 + (k_{z} + eA_z)^2},\nonumber\\
\alpha &=& \frac{eA_{\phi}}{q} = -\frac{\Phi_{\phi}}{\Phi_0}.
\label{const}
\end{eqnarray}
where $\Phi_0=\frac{2\pi}{e}$, the quantum flux. In \eqref{Solu1} $J_\mu(z)$ represents the Bessel function \cite{Abra}. 

The quasiperiodicity condition \eqref{QPC} provides a discretization  of the quantum number $k_z$ as shown below:
\begin{equation}
k_z = k_l = \frac{2\pi}{L}(l + \beta), \qquad \text{with}\qquad l = 0,\pm 1, \pm2,...\;.
\label{momentum}
\end{equation}
Therefore 
\begin{eqnarray}
E=E_{l} = \sqrt{\lambda^2 + p^2 + \vec{k}^2 + \tilde{k}_l^2},
\label{const2}
\end{eqnarray}
where 
\begin{eqnarray}
\tilde{k}_l &=& \frac{2\pi}{L}(l + \tilde{\beta}),\nonumber\\
\tilde{\beta} &=& \beta + \frac{eA_zL}{2\pi} = \beta - \frac{\Phi_z}{\Phi_0}.
\label{const3}
\end{eqnarray}

The constant $C$ in \eqref{Solu1} can be obtained by the normalization condition below, 
\begin{eqnarray}
\int d^Dx\sqrt{|g|}g^{00}\varphi_{\sigma'}^{*}(x)\varphi_{\sigma}(x)= \frac{1}{2E}\delta_{\sigma,\sigma'}  \   ,
\label{NC}
\end{eqnarray}
where the delta symbol on the right-hand side is understood as Dirac delta 
function for the continuous quantum number, $\lambda$, $p$ and ${\vec{k}}$, and
Kronecker delta for the discrete ones, $n$ and $k_l$. From \eqref{NC} one finds 
\begin{eqnarray}
|C|= \sqrt{\frac{qa^{1-D}\lambda p}{2E (2\pi)^{D-3}L}}.
\label{NC3}
\end{eqnarray}
So, the normalized bosonic wave-function reads,
\begin{equation}
\varphi_{\sigma}(x) =  \sqrt{\frac{qa^{1-D}\lambda p}{2E (2\pi)^{D-3}L}}w^{\frac{D}{2}}J_{\nu}(pw)J_{q|n +\alpha|}(\lambda r)e^{-iE_{l} t + iqn\phi + ik_{l}z + i\vec{k}\cdot\vec{x}_{\parallel}}  \  .
\label{COS}
\end{equation}

The properties of the vacuum state can be given by the positive 
frequency Wightman function, $W(x,x')=\left\langle 0|\hat{\varphi}(x) \hat{\varphi}^{*}(x')|0 \right\rangle$, where $|0 \rangle$ stands for the vacuum state. To evaluate it we use the mode sum positive frequency Wightman function
\begin{equation}
W(x,x') = \sum_{\sigma}\varphi_{\sigma}(x)\varphi_{\sigma}^{*}(x'),
\label{wight}
\end{equation}
where $\sum_{\sigma}$ contains summation over discrete quantum numbers and integration over continuous ones. 
Substituting \eqref{COS} into \eqref{wight} we obtain,
\begin{eqnarray}
W(x,x') = \frac{qa^{1-D}(ww')^{\frac{D}{2}}}{2(2\pi)^{D-3}L}\sum_{n=-\infty}^{\infty}e^{inq\Delta\phi}\sum_{l=-\infty}^{\infty}\int d\vec{k}\int_0^{\infty}dpp\int_0^{\infty}d\lambda\lambda\nonumber\\
 \times J_{q|n +\alpha|}(\lambda r)J_{q|n +\alpha|}(\lambda r')J_{\nu}(pw)J_{\nu}(pw')\frac{e^{-iE\Delta t+ ik_{l}\Delta z + i\vec{k}\cdot\Delta\vec{r}_{\parallel}}}{E}
\label{wight2}
\end{eqnarray}
where $\Delta t=t-t', \Delta \phi=\phi-\phi', \Delta z=z-z'$ and $\Delta \vec{x}_{\parallel}= \vec{x}_{\parallel}-\vec{x}_{\parallel}'$.

In order to develop the summation over the quantum number $l$ we shall apply
the Abel-Plana summation formula \cite{Saharian2010}, which is given by
\begin{eqnarray}
\sum_{l=-\infty}^{\infty}g(l+\tilde{\beta})f(|l+\tilde
{\beta}|)&=&\int_{0}^{\infty}du[g(u)+g(-u)]f(u)\nonumber\\
&+&i\int_{0}^{\infty}du[f(iu)-f(-iu)]\sum_{j=\pm1}^{}\frac{g(i j u)}{e^{2\pi(u+i j\tilde{\beta})}-1}  \  .
\label{Abel-Plana}
\end{eqnarray}
For this case, we can identify
\begin{eqnarray}
g(u)&=&e^{2\pi i u\Delta z/L} \nonumber\\
f(u)&=&\frac{e^{-i\Delta t\sqrt{\lambda^2+p^2+\vec{k}^2+(2\pi u/L)^2}}}{\sqrt{\lambda^2+p^2+\vec{k}^2+(2\pi u/L)^2}} \  , 
\end{eqnarray}

Using \eqref{Abel-Plana}, we can write the Wightman function as
\begin{equation}
W(x,x')=W_{cs}(x,x')+W_{c}(x,x'),
\label{propagator}
\end{equation}
where the first term is a contribution due to the cosmic string spacetime in AdS bulk, and the second one is due to the compactification. Both terms in the right hand side of the above equation has been individually developed in \cite{Santos:2018ttf}. Explicitly, they may be written as
\begin{eqnarray}
W_{cs}(x,x')&=&\frac{q(ww')^{\frac{D}{2}}e^{ -ieA_z\Delta z}}{(2\pi)^{D-2}a^{D-1}}\int d\vec{k}e^{i\vec{k}\cdot\Delta\vec{x}_{\parallel}}\int_0^{\infty}dppJ_{\nu}(pw)J_{\nu}(pw')\nonumber\\
&\times&\sum_{n=-\infty}^{\infty}e^{inq\Delta\phi}\int_0^{\infty}d\lambda\lambda J_{q|n +\alpha|}(\lambda r)J_{q|n +\alpha|}(\lambda r')\nonumber\\&\times&\int_{0}^{\infty}\frac{ds}{s}e^{-s^2(\lambda^2+p^2+\vec{k}^2)-(\Delta z^2-\Delta t^2)/4s^2},
\label{propagator-cs-2}
\end{eqnarray}
and
\begin{eqnarray}
W_{c}(x,x')&=&\frac{q(ww')^{\frac{D}{2}}e^{-ieA_z\Delta z}}{(2\pi)^{D-2}a^{D-1}}\int d\vec{k}e^{i\vec{k}\cdot\Delta\vec{x}_{\parallel}}\int_0^{\infty}dppJ_{\nu}(pw)J_{\nu}(pw')\nonumber\\
&\times&\sum_{n=-\infty}^{\infty}e^{inq\Delta\phi}\int_0^{\infty}d\lambda\lambda J_{q|n +\alpha|}(\lambda r)J_{q|n +\alpha|}(\lambda r')\nonumber\\
&\times&	\sum_{j=\pm 1}^{}\sum_{l=1}^{\infty}e^{-2\pi i\tilde{\beta}jl}\int_{0}^{\infty}\frac{ds}{s}e^{-s^2(\lambda^2+p^2+\vec{k}^2)-[(lL+j\Delta z)^2-\Delta t^2]/4s^2}  \  .
\label{propagator-compactification-2}
\end{eqnarray}
Substituting \eqref{propagator-cs-2} and \eqref{propagator-compactification-2} into \eqref{propagator}, and after long but straightforward steps, we get
\begin{eqnarray}
W(x,x')&=&\frac{qe^{-ieA_z\Delta z}}{2(2\pi)^{\frac{D}{2}}a^{D-1}}\bigg(\frac{ww'}{rr'}\bigg)^{\frac{D}{2}}\sum_{l=-\infty}^{\infty}e^{-2\pi i\tilde{\beta}l}\int_{0}^{\infty}d\chi \chi^{\frac{D}{2}-1}e^{-\chi u_{l}^{2}/2rr'}
I_{\nu}\bigg(\frac{ww'}{rr'}\chi\bigg)\nonumber\\
&\times&\sum_{n=-\infty}^{\infty}e^{iqn\Delta\phi}I_{q|n+\alpha|}(\chi)  \  , 
\label{propagator-to-sum}
\end{eqnarray}
where we have introduced a new variable $\chi=rr'/2s^2$, and defined
\begin{equation}
u_{l}^{2}=r^2+r'^2+w^2+w'^2+(lL+\Delta z)^2+\Delta \vec{x}^{2}_{\parallel}-\Delta t^2  \ .
\end{equation}
The parameter $\alpha$ in Eq.\eqref{const} is written in the form
\begin{equation}
\alpha=n_{0}+\varepsilon, \ \textrm{with}\ |\varepsilon|<\frac{1}{2},
\label{const-2}
\end{equation}
being $n_{0}$ an integer number. This allow us to sum over the quantum number $n$ in \eqref{propagator-to-sum}, using the result obtained in \cite{deMello:2014ksa}, given below,
\begin{eqnarray}
&&\sum_{n=-\infty}^{\infty}e^{iqn\Delta\phi}I_{q|n+\alpha|}(\chi)=\frac{1}{q}\sum_{k}e^{\chi\cos(2\pi k/q-\Delta\phi)}e^{i\alpha(2\pi k -q\Delta\phi)}\nonumber\\
&-&\frac{e^{-iqn_{0}\Delta\phi}}{2\pi i}\sum_{j=\pm1}je^{ji\pi q|\varepsilon|}
\int_{0}^{\infty}dy\frac{\cosh{[qy(1-|\varepsilon|)]}-\cosh{(|\varepsilon| qy)e^{-iq(\Delta\phi+j\pi)}}}{e^{\chi\cosh{(y)}}\big[\cosh{(qy)}-\cos{(q(\Delta\phi+j\pi))}\big]},
\label{summation-formula}
\end{eqnarray}
where
\begin{equation}
-\frac{q}{2}+\frac{\Delta\phi}{\Phi_{0}}\le k\le \frac{q}{2}+\frac{\Delta\phi}{\Phi_{0}}  \   .
\end{equation}
Thus, the substitution of \eqref{summation-formula} into \eqref{propagator-to-sum}, allow us to integrate over $\chi$ with the help of \cite{Grad}, yielding
\begin{eqnarray}
W(x,x')&=&\frac{e^{-ieA_z\Delta z}}{(2\pi)^{\frac{D+1}{2}}a^{D-1}}\sum_{l=-\infty}^{\infty}e^{-2\pi i\tilde{\beta}l}\Bigg\{\sum_{k}e^{i\alpha(2\pi k-q\Delta\phi)}F_{\nu-1/2}^{(D-1)/2}({u}_{lk})\nonumber\\
&-&q\frac{e^{-iqn_{0}\Delta\phi}}{2\pi i}\sum_{j=\pm1}je^{ji\pi q|\varepsilon|}
\int_{0}^{\infty}dy\frac{\cosh{[(1-|\varepsilon|)qy]}-\cosh{(|\varepsilon|q y)e^{-iq(\Delta\phi+j\pi)}}}{\cosh{(qy)}-\cos{(q(\Delta\phi+j\pi))}}\nonumber\\
&\times&F_{\nu-1/2}^{(D-1)/2}({u}_{ly})\Bigg\},
\label{full-propagator}
\end{eqnarray}
where we have introduced the notation
\begin{eqnarray}
F_{\gamma}^{\mu}(u)&=&e^{-i\pi\mu}\frac{Q^{\mu}_{\gamma}(u)}{(u^2-1)^{\mu/2}}  \nonumber\\  
&=&\frac{\sqrt{\pi}\Gamma(\gamma+\mu+1)}{2^{\gamma+1}\Gamma(\gamma+3/2)u^{\gamma+\mu+1}}F\bigg(\frac{\gamma+\mu}{2}+1,\frac{\gamma+\mu+1}{2};\gamma+\frac{3}{2};\frac{1}{u^{2}}\bigg).
\label{function-2}
\end{eqnarray}
being $Q_{\gamma}^{\mu}(u)$ the associated Legendre function of second kind and $F(a,b;c;z)$ the hypergeometric function \cite{Abra}. In \eqref{full-propagator}, the arguments of the function $F_{\gamma}^{\mu}$ are given by
\begin{eqnarray}
u_{lk}&=&1+\frac{r^2+r'^2-2rr'\cos{(2\pi k/q-\Delta\phi)}+\Delta w^2+(lL+\Delta z)^2+\Delta\vec{x}^{2}_{\parallel}-\Delta t^2}{2ww'}\nonumber\\
u_{ly}&=&1+\frac{r^2+r'^2+2rr'\cosh{(y)}+\Delta w^2+(lL+\Delta z)^2+\Delta\vec{x}^{2}_{\parallel}-\Delta t^2}{2ww'}.
\end{eqnarray}
The Wightman function can be written as a sum of different components,
\begin{equation}
	W(x,x')=W_{\rm{AdS}}(x,x')+W_{\rm{cs}}(x,x')+W_{\rm{c}}(x,x'),
	\label{wightman-function-expanded}
\end{equation}
where the first term is a pure AdS spacetime contribution ($k=0$ and $l=0$), the second one is due to the cosmic string ($k\neq0$ and $l=0$) and the third term is associated with the compactification ($l\neq0$).
\section{The VEV of the Field Squared}
\label{sec3}
 The VEV of the field squared is formally obtained from the Wightman function by taking the coincidence limit, as shown below:
\begin{equation}
	\langle|\varphi|^2\rangle=\lim\limits_{x'\rightarrow x}W(x,x') \  .
\end{equation}
By substituting \eqref{wightman-function-expanded} into the above expression, we see that $\langle|\varphi|^2\rangle$ presents three contributions: 
\begin{equation}
\langle|\varphi|^2\rangle=\langle|\varphi|^2\rangle_{\rm{AdS}}+\langle|\varphi|^2\rangle_{\rm{cs}}+\langle|\varphi|^2\rangle_{\rm{c}}.
\end{equation}
However the above expression provides a divergent result. Because the presence of a cosmic string does not introduce additional curvature for points outside of the string's core, the divergence comes only from the contribution due to the pure AdS spacetime. The analysis of the renormalized VEV of the field squared in AdS bulk has been developed in the literature \cite{Burgess}-\cite{Caldarelli}. In this paper we will concentrate in the contributions induced by the string and the compactification. 

Considering the cosmic string component of the Wightman function in \eqref{full-propagator} and taking the coincidence limit, we have 
\begin{eqnarray}
	\langle|\varphi|^2\rangle_{\rm{cs}}&=&\frac{2}{(2\pi)^{\frac{D+1}{2}}a^{D-1}}\Bigg[\sideset{}{'}\sum_{k=1}^{[q/2]}\cos{(2\pi k\varepsilon)}F^{(D-1)/2}_{\nu-1/2}(v_{k})\nonumber\\
	&-&\frac{q}{\pi}\int_{0}^{\infty}dy\frac{f(q,\varepsilon,2y)}{\cosh(2qy)-\cos(q\pi)}F^{(D-1)/2}_{\nu-1/2}(v_{y})\Bigg]  \  ,
	\label{phi-squared}
\end{eqnarray}
where we have introduced the following notations:
\begin{eqnarray}
v_{k}&=&1+2(r/w)^2\sin^2{(\pi k/q)}, \nonumber\\
v_{y}&=&1+2(r/w)^2\cosh^2{(y)}  \  ,
\end{eqnarray}
being,
\begin{equation}
	f(q,\varepsilon,2y)=\sin(|\varepsilon|q\pi)\cosh((1-|\varepsilon|)2qy)+\cosh(2|\varepsilon|qy)\sin((1-|\varepsilon|)q\pi) \ .
	\label{ffunction}
\end{equation}
In Eq.\eqref{phi-squared} $[q/2]$ represents the integer part of $q/2$ and the prime on the sign of the summation means that in the case $q=2p$ the term $1/2$ should be taken with the coefficient $1/2$.

Some interesting asymptotic behaviors for Eq.\eqref{phi-squared} are presented below. We start considering $r/w\rightarrow0$. We can use the asymptotic formula for the hypergeometric function for small arguments \cite{Abra} to rewrite Eq.\eqref{phi-squared} as
\begin{equation}
	\langle|\varphi|^2\rangle_{cs}\approx\frac{2\Gamma\big(\frac{D-1}{2}\big)}{(4\pi)^{\frac{D+1}{2}}}\bigg(\frac{w}{ar}\bigg)^{D-1}\bigg[\sideset{}{'}\sum_{k=1}^{[q/2]}\frac{\cos(2\pi k\varepsilon)}{\sin^{D-1}(\pi k/q)}-\frac{q}{\pi}\int_{0}^{\infty}dy\frac{f(q,\varepsilon,2y)\cosh^{1-D}(y)}{\cosh(2qy)-\cos(q\pi)}\bigg] \ .
	\label{field-squared-cs-assymp}
\end{equation}
We can notice that the above result diverges with inverse of proper distance from the string at the power $(D-1)$. In other words, for a fixed value of $w$, $\langle|\varphi|^2\rangle_{cs}$ goes to infinity  for points near the string.

On the other hand, for $w\ll r$ and by using the asymptotic expression below \cite{deMello:2014hya},
\begin{equation}
	F_{\nu-1/2}^{(D-1)/2}(u)\approx\frac{\sqrt{\pi}\Gamma(D/2+\nu)}{2^{\nu+1/2}\Gamma(\nu+1)u^{D/2+\nu}} \ ,
\end{equation}
we get
\begin{eqnarray}
	\langle|\varphi|^2\rangle_{cs}&\approx&\frac{2^{-2\nu}\Gamma(D/2+\nu)}{(4\pi)^{\frac{D}{2}}\Gamma(\nu+1)a^{D-1}}\bigg(\frac{w}{r}\bigg)^{D+2\nu}\bigg[\sideset{}{'}\sum_{k=1}^{[q/2]}\frac{\cos(2\pi k\varepsilon)}{\sin^{D+2\nu}(\pi k/q)}\nonumber\\
	&-&\frac{q}{\pi}\int_{0}^{\infty}dy\frac{f(q,\varepsilon,2y)\cosh^{-D-2\nu}(y)}{\cosh(2qy)-\cos(q\pi)}\bigg] \ .
	\label{field-squared-cs-assymp2}
\end{eqnarray}
From the above expression, we observe that for fixed values of the radial coordinate $r$, the string-induced contribution goes to zero near the AdS boundary as $w^{D+2\nu}$.

For $\nu\gg1$, we have
\begin{eqnarray}
	\langle|\varphi|^2\rangle_{cs}&\approx&\frac{\nu^{\frac{D}{2}-1}}{(2\pi)^{\frac{D}{2}}a^{D-1}}\Bigg[\sideset{}{'}\sum_{k=1}^{[q/2]}\cos(2\pi k\varepsilon)\frac{\big(v_{k}+\sqrt{v_{k}^{2}-1}\big)^{-\nu}}{(v_{k}^{2}-1)^{\frac{D}{4}}}\nonumber\\
	&-&\frac{q}{\pi}\int_{0}^{\infty}dy\frac{f(q,\varepsilon,2y)}{\cosh(2qy)-\cos(q\pi)}\frac{\big(v_{y}+\sqrt{v_{y}^{2}-1}\big)^{-\nu}}{(v_{y}^{2}-1)^{\frac{D}{4}}}\Bigg].
	\label{field-squared-cs-assymp3}
\end{eqnarray}

For a conformally coupled massless scalar field we have $\nu=1/2$, and by expressing the associated Legendre function in terms of  hypergeometric function \cite{Abra,Grad}, we can  write a more convenient expression for $F_{\nu-1/2}^{(D-1)/2}(u)$ \cite{deMello:2014hya}, given by
\begin{eqnarray}
F_{0}^{(D-1)/2}(u)=-\frac{\Gamma\big(\frac{D-1}{2}\big)}{2}\bigg[(1+u)^{-(D-1)/2}-(u-1)^{-(D-1)/2}\bigg].
\label{function-3}
\end{eqnarray} 
Substituting \eqref{function-3} into \eqref{phi-squared}, we obtain
\begin{equation}
\langle|\varphi|^2\rangle_{\rm{cs}}=\bigg(\frac{w}{a}\bigg)^{D-1}\bigg[\langle|\phi|^2\rangle_{\rm{cs}}^{\rm{(M)}}+\langle|\phi|^2\rangle_{\rm{cs,b}}^{\rm{(M)}}\bigg],
\label{conformal-phi-squared}
\end{equation}
where
\begin{equation}
\langle|\varphi|^2\rangle_{\rm{cs}}^{\rm{(M)}}=\frac{2\Gamma(\frac{D-1}{2})}{(4\pi)^{\frac{D+1}{2}}r^{D-1}}\Bigg[\sideset{}{'}\sum_{k=1}^{[q/2]}\frac{\cos{(2\pi k\varepsilon)}}{\sin^{D-1}(\pi k/q)}-\frac{q}{\pi}\int_{0}^{\infty}dy\frac{f(q,\varepsilon,2y)\cosh^{1-D}(y)}{\cosh(2qy)-\cos(q\pi)}\Bigg]
\label{conformal-sf-cs}
\end{equation}
is the VEV for the boundary-free cosmic string geometry corrected by the presence of a magnetic flux running through the string's core.
The second term in \eqref{conformal-phi-squared},
\begin{eqnarray}
\langle|\varphi|^2\rangle_{\rm{cs,b}}^{\rm{(M)}}&=&-\frac{2\Gamma(\frac{D-1}{2})}{(4\pi)^{\frac{D+1}{2}}}\Bigg[\sideset{}{'}\sum_{k=1}^{[q/2]}\cos(2\pi k\varepsilon)\bigg(w^2+r^2\sin^2(\pi k/q)\bigg)^{-\frac{(D-1)}{2}}\nonumber\\
&-&
\frac{q}{\pi}\int_{0}^{\infty}dy\frac{f(q,\varepsilon,2y)}{\cosh(2qy)-\cos(q\pi)}\bigg(w^2+r^2\cosh^2(y)\bigg)^{-\frac{(D-1)}{2}}\Bigg],
\end{eqnarray}
is the contribution induced by the boundary located at $w=0$. It is finite at the string's core for $w\neq 0$. In addition, for $r\gg w$ this contribution tends to cancel \eqref{conformal-sf-cs} inside the square bracket in Eq.\eqref{conformal-phi-squared}.

\begin{figure}[!htb]
	\begin{center}
		\centering
		\includegraphics[scale=0.4]{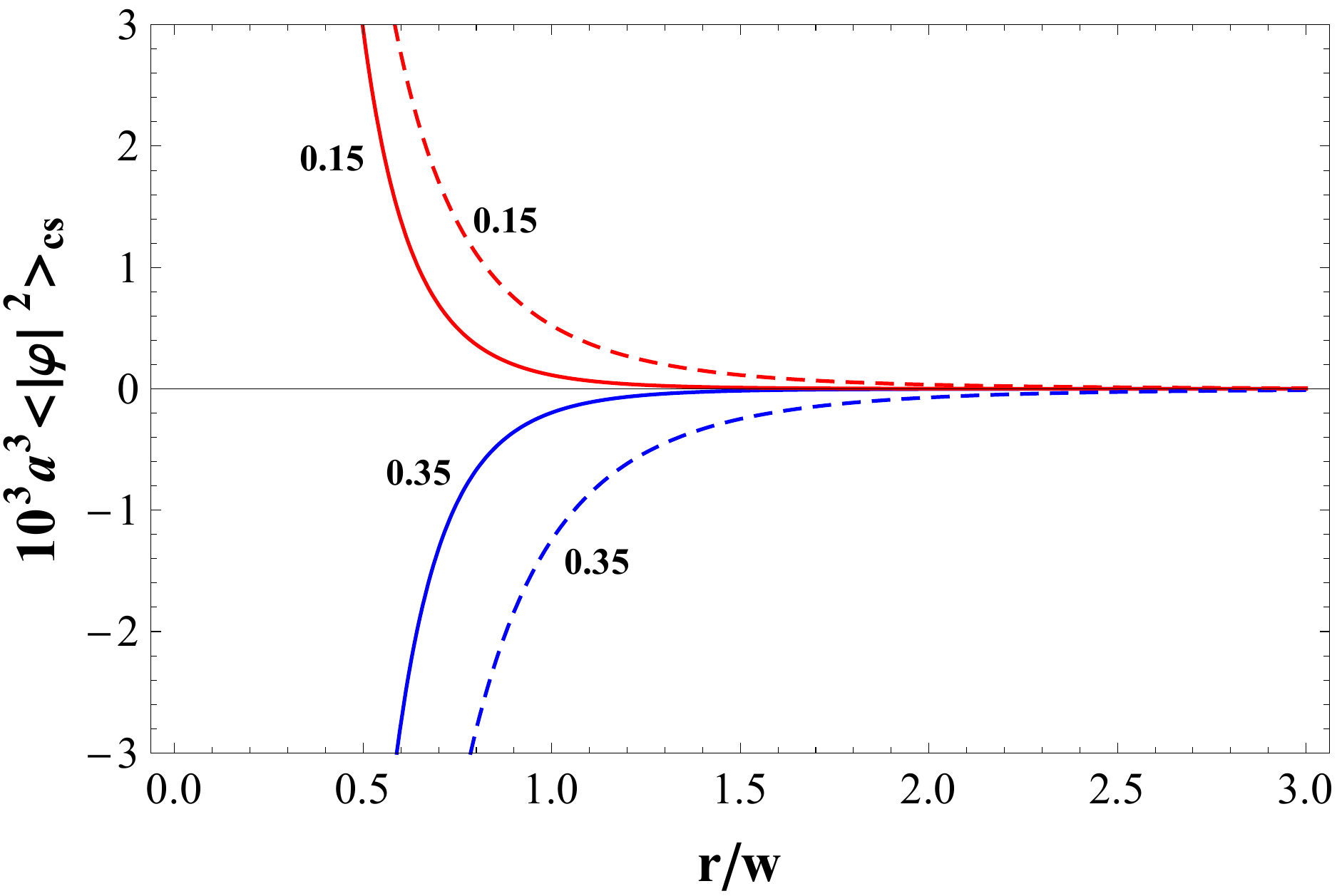}
		\caption{The VEV of the field squared without compactification for $D=4$ in Eq.\eqref{phi-squared} is plotted in units of $a^{3}$, in terms of $r/w$, for $q=2.5$ and $m=0$. The numbers near the curves correspond to the values of the parameter $\varepsilon$. The full lines correspond to the massless minimum coupling field while the dashed ones correspond to the conformal massless case.}
		\label{fig1}
	\end{center}
\end{figure}

In Fig.\ref{fig1} we exhibit the behavior of the VEV of the field squared associated with the string,  as a function of $r/w$, that is the proper distance from the string in units of AdS curvature radius. We note from this figure that the parameters associated with magnetic flux along the string, $\varepsilon$, and the curvature coupling, $\xi$, can change the intensity and behavior of the field squared. 

Now considering the  component $l\neq0$ in \eqref{full-propagator} and taking the coincidence limit, we get
\begin{equation}
	\langle|\varphi|^2\rangle_{\rm{c}}=\langle|\varphi|^2\rangle_{\rm{c}}^{(0)}+\langle|\varphi|^2\rangle_{\rm{c}}^{(q,\varepsilon)} \ ,
	\label{field-squared-c}
\end{equation}
where the first term on the right-hand side of the above expression is
\begin{equation}
	\langle|\varphi|^2\rangle_{\rm{c}}^{(0)}=\frac{2}{(2\pi)^{\frac{D+1}{2}}a^{D-1}}\sum_{l=1}^{\infty}\cos(2\pi \tilde{\beta}l)F^{(D-1)/2}_{\nu-1/2}(v_{l0})) \ ,
	\label{term-k-0}
\end{equation}
with $v_{l0}$ given in \eqref{arg-phi-c} for $k=0$. This contribution does not depend on $q$ and $\varepsilon$. For values of $v_{l0}$ close to the unit, the above expression has the assymptotic form
\begin{equation}
	\langle|\varphi|^2\rangle_{\rm{c}}^{(0)}\approx\frac{\Gamma(\frac{D-1}{2})}{2\pi^{\frac{D+1}{2}}a^{D-1}}\bigg(\frac{w}{L}\bigg)^{D-1}\sum_{l=1}^{\infty}\frac{\cos(2\pi \tilde{\beta}l)}{l^{D-1}} \ .
	\label{assymp-c-0}
\end{equation}
In the case of a conformally coupled massless scalar field, we have
\begin{equation}
	\langle|\varphi|^2\rangle_{\rm{c}}^{(0)}=\frac{\Gamma(\frac{D-1}{2})}{2\pi^{\frac{D+1}{2}}a^{D-1}}\bigg(\frac{w}{L}\bigg)^{D-1}\sum_{l=1}^{\infty}\cos(2\pi \tilde{\beta}l)\Bigg\{\frac{1}{l^{D-1}}-\frac{1}{\big[l^2+\big(\frac{2w}{L}\big)^2\big]^{\frac{D-1}{2}}}\Bigg\} \ .
	\label{conform}
\end{equation}
We note that the above expression goes to zero near the AdS boundary. On the other hand for points near the AdS horizon, \eqref{conform} coincides with the one given in \eqref{assymp-c-0}. 

As to the second term in \eqref{field-squared-c}, we can see that it depends on the magnetic fluxes and the parameter
associated with the cosmic string, $q$, and it is given by
\begin{eqnarray}
	\langle|\varphi|^2\rangle_{\rm{c}}^{(q,\varepsilon)}&=&\frac{4}{(2\pi)^{\frac{D+1}{2}}a^{D-1}}\sum_{l=1}^{\infty}\cos(2\pi \tilde{\beta}l)\Bigg[\sideset{}{'}\sum_{k=1}^{[q/2]}\cos{(2\pi k\varepsilon)}F^{(D-1)/2}_{\nu-1/2}(v_{lk})\nonumber\\
	&-&\frac{q}{\pi}\int_{0}^{\infty}dy\frac{f(q,\varepsilon,2y)}{\cosh(2qy)-\cos(q\pi)}F^{(D-1)/2}_{\nu-1/2}(v_{ly})\Bigg] \ ,
	\label{VEV-field-squared-c}
\end{eqnarray}
where
\begin{eqnarray}
v_{lk}&=&1+\frac{(lL)^2+4r^2\sin^2{(\pi k/q)}}{2w^2}\nonumber\\
v_{ly}&=&1+\frac{(lL)^2+4r^2\cosh^2{(y)}}{2w^2}  \  .
\label{arg-phi-c}
\end{eqnarray}
\begin{figure}[!htb]
	\begin{center}
		\includegraphics[scale=0.4]{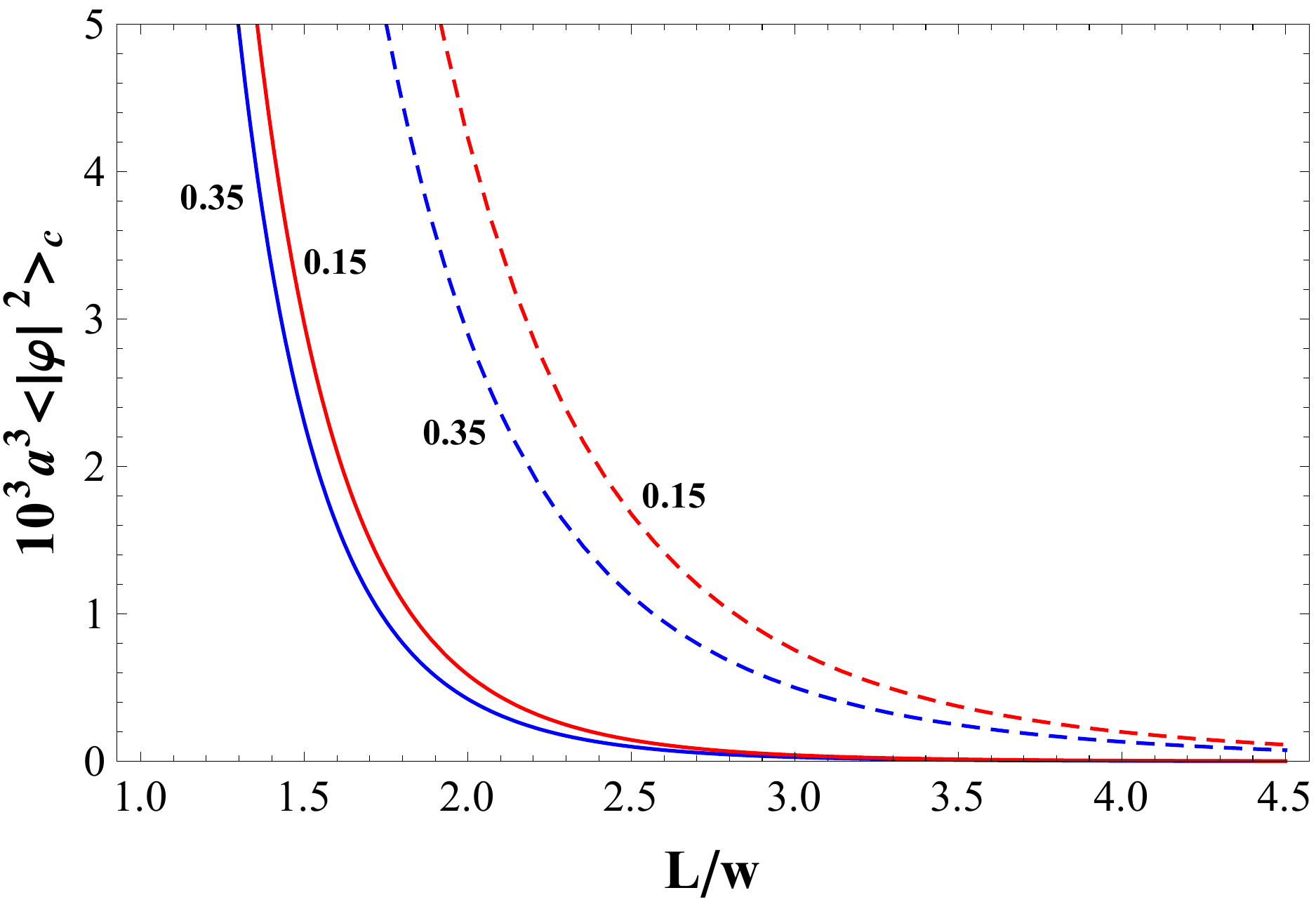}
		\quad
		\includegraphics[scale=0.4]{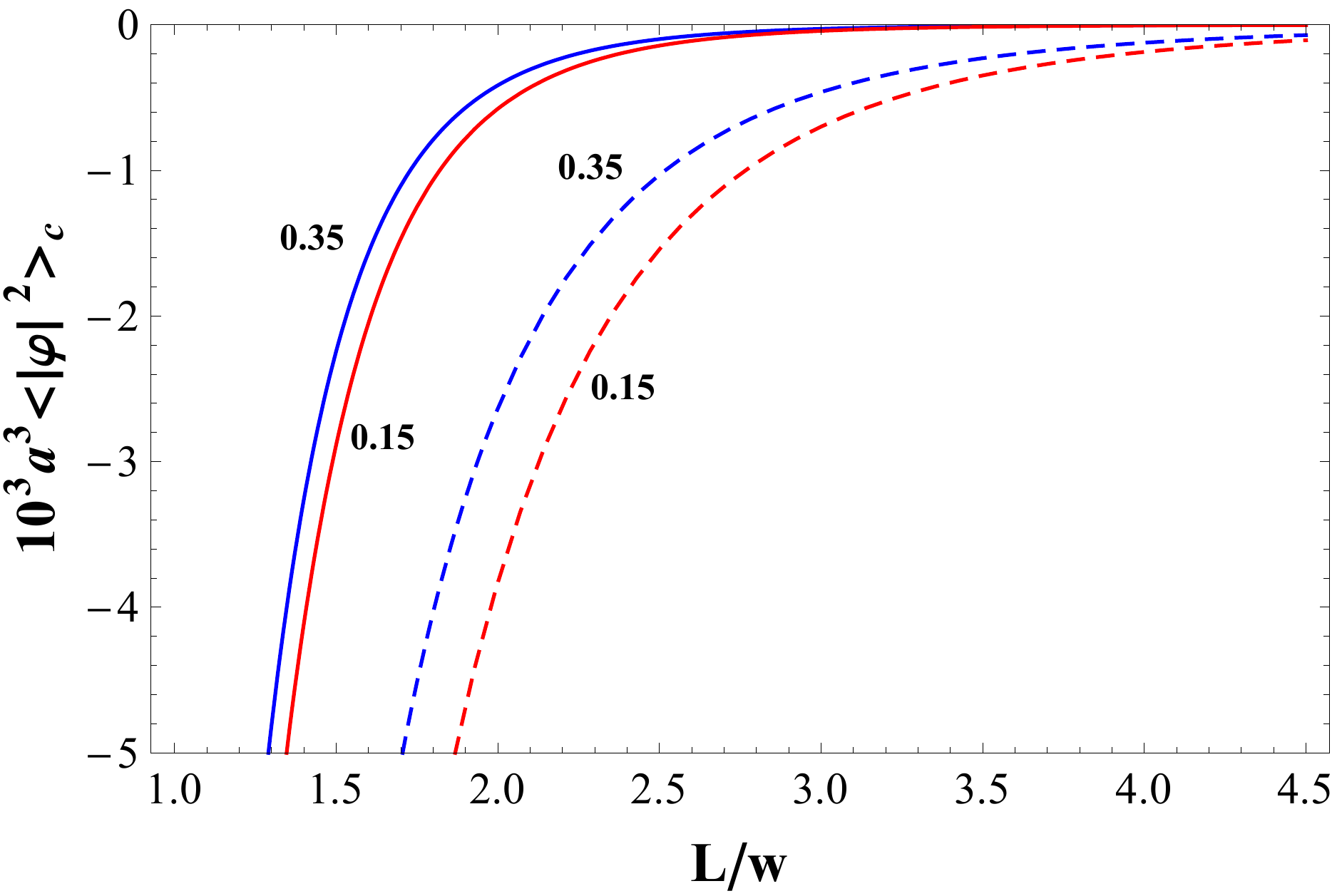}
		\caption{The VEV of the field squared induced by the compactification in Eq.\eqref{field-squared-c} is plotted for $D=4$, in units of $a^{3}$, as a function of $L/w$,  for $q=2.5$, $r/w=0.5$ and $m=0$. The numbers near the curves correspond to the values of parameter $\varepsilon$. The left panel corresponds to $\tilde{\beta}=0$ and the right one to $\tilde{\beta}=0.5$. The full lines correspond to the massless minimum coupling field while the dashed ones correspond to the conformal massless case.}
		\label{fig2}
	\end{center}
\end{figure}

In Fig.\ref{fig2} we present the behavior of \eqref{field-squared-c} as function of $L/W$, considering different values of $\varepsilon$. In the left panel we assume $\tilde{\beta}=0$, and in the right panel $\tilde{\beta}=0.5$, that correspond the untwisted and twisted field cases, respectively. We note from these figures that the VEV of the field squared depends strongly with the parameter $\tilde{\beta}$. 

We can notice that $\langle |\varphi|^2\rangle_{c}^{(q,\varepsilon)}$ is finite for $r=0$.
In the regime $L/w\gg1$, Eq.\eqref{VEV-field-squared-c} assumes the following asymptotic form
\begin{eqnarray}
\langle |\varphi|^2\rangle_{c}^{(q,\varepsilon)}&\approx& \frac{2^{1-2\nu}\Gamma(D/2+\nu)}{(4\pi)^{\frac{D}{2}}\Gamma(\nu+1)a^{D-1}}\bigg(\frac{w}{L}\bigg)^{D+2\nu}\sum_{l=1}^{\infty}\cos(2\pi l\tilde{\beta})\nonumber\\
&\times&\Bigg\{\sideset{}{'}\sum_{k=1}^{[q/2]}\cos(2\pi k\varepsilon)
\bigg[\frac{l^2}{4}+\bigg(\frac{r}{L}\bigg)^{2}\sin^2(\pi k/q)\bigg]^{-\frac{D}{2}-\nu}
\nonumber\\
&-&\frac{q}{\pi}\int_{0}^{\infty}dy\frac{f(q,\varepsilon,2y)}{\cosh(2qy)-\cos(\pi q)}\bigg[\frac{l^2}{4}+\bigg(\frac{r}{L}\bigg)^{2}\cosh^2(y)\bigg]^{-\frac{D}{2}-\nu}\Bigg\} \ .
\label{assymp-field-squared-c}
\end{eqnarray}
For a conformally coupled massless scalar field, we have
\begin{eqnarray}
	\langle|\varphi|^2\rangle_{\rm{c}}^{(q,\varepsilon)}&=&\frac{4\Gamma\big(\frac{D-1}{2}\big)}{(4\pi)^{\frac{D+1}{2}}a^{D-1}}\sum_{l=1}^{\infty}\cos(2\pi \tilde{\beta}l)\Bigg\{\sideset{}{'}\sum_{k=1}^{[q/2]}\cos(2\pi k\varepsilon)\Bigg[\bigg(\frac{r^2}{w^2}\sin^2(\pi k/q)+\frac{(lL)^2}{4w^2}\bigg)^{-\frac{D-1}{2}}\nonumber\\
	&-&\bigg(1+\frac{r^2}{w^2}\sin^2(\pi k/q)+\frac{(lL)^2}{4w^2}\bigg)^{-\frac{D-1}{2}}\Bigg]-\frac{q}{\pi}\int_{0}^{\infty}dy\frac{f(q,\varepsilon,2y)}{\cosh(2qy)-\cos(q\pi)}\nonumber\\
	&\times&\Bigg[\bigg(\frac{r^2}{w^2}\cosh^2(y)+\frac{(lL)^2}{4w^2}\bigg)^{-\frac{D-1}{2}}-\bigg(1+\frac{r^2}{w^2}\cosh^2(y)+\frac{(lL)^2}{4w^2}\bigg)^{-\frac{D-1}{2}}\Bigg]\Bigg\} \ .
	\label{field-squared-c-conformal}
\end{eqnarray}
Thus, we can see that the above equation is expressed in terms of elementary functions.
\section{VEV of the Energy-Momentum Tensor}
\label{sec4}
Having obtained the Wightman function and the mean field square, we are now in position to calculate the VEV of the energy-momentum tensor by making use of the formula developed in the Appendix \ref{Apendx},
\begin{equation}
	\langle T_{\mu\nu}\rangle=(D_{\mu}D_{\nu'}^{\dagger}+D_{\mu'}^{\dagger}D_{\nu})W(x,x')-2[\xi R_{\mu\nu}+\xi\nabla_{\mu}\nabla_{\nu}-(\xi-1/4)g_{\mu\nu}\nabla_{\alpha}\nabla^{\alpha}]]\langle|\varphi|^2\rangle,
	\label{E-M-Tensor-formula}
\end{equation}
where $R_{\mu\nu}=-Dg_{\mu\nu}/a^2$ is the Ricci tensor for the AdS spacetime and $D_{\mu}=\nabla_{\mu}+ieA_{\mu}$. Similarly to the VEV of the field squared, the VEV of the energy-momentum tensor can be decomposed as
\begin{equation}
	\langle T_{\mu\nu}\rangle=\langle T_{\mu\nu}\rangle_{\rm{AdS}}+\langle T_{\mu\nu}\rangle_{\rm{cs}}+\langle T_{\mu\nu}\rangle_{\rm{c}} \  .
	\label{TEM0}
\end{equation}
As consequence of maximal symmetry of AdS spacetime and the vacuum state under consideration, one has $\langle T_{\mu\nu}\rangle_{\rm{AdS}}=\text{const}\cdot g_{\mu\nu}$ \cite{Caldarelli}. Therefore, the corresponding VEV of the energy-momentum tensor in pure AdS spacetime is completely determined by its trace. In this paper, we are mainly interested in the string and compactified induced parts. 
For convenience, we will work each term separately.

The covariant d'Alembertian acting in the quantities given in \eqref{phi-squared} and \eqref{field-squared-c}, respectively, gives
\begin{eqnarray}
	\Box\langle |\varphi|^2\rangle_{\rm{cs}}&=&-\frac{8}{(2\pi)^{\frac{D+1}{2}}a^{D+1}}\Bigg[\sideset{}{'}\sum_{k=1}^{[q/2]}\cos(2\pi k\varepsilon)g(v_{k},\sin(\pi k/q))\\\nonumber
	&-&\frac{q}{\pi}\int_{0}^{\infty}dy\frac{f(q,\varepsilon,2y)g(v_{y},\cosh(y))}{\cosh(2qy)-\cos(q\pi)}\Bigg]
\end{eqnarray}
and
\begin{eqnarray}
	\Box\langle|\varphi|^2\rangle_{\rm{c}}&=&-\frac{16}{(2\pi)^{\frac{D+1}{2}}a^{D+1}}\sum_{l=1}^{\infty}\cos(2\pi\tilde{\beta}l)\Bigg\{\sideset{}{'_*}\sum_{k=0}^{[q/2]}\cos{(2\pi k\varepsilon)}\Bigg[g(v_{lk},\sin(k\pi/q))-\frac{(lL\sin(\pi k/q))^2}{w^2}\nonumber\\&\times&\frac{d^2}{dv_{lk}^2}F^{(D-1)/2}_{\nu-1/2}(v_{lk})\Bigg]
	-\frac{q}{\pi}\int_{0}^{\infty}\frac{dyf(q,\varepsilon,2y)}{\cosh(2qy)-\cos(q\pi)}\Bigg[g(v_{ly},\cosh(y))-\frac{(lL\cosh(y))^2}{w^2}\nonumber\\
	&\times&\frac{d^2}{dv_{ly}^2}F^{(D-1)/2}_{\nu-1/2}(v_{ly})\Bigg]\Bigg\},
\end{eqnarray}
with the notation
\begin{equation}
	g(u,v)=(u-1)(2v^2+u-1)\frac{d^2}{du^2}F^{(D-1)/2}_{\nu-1/2}(u)+\bigg[2v^2+\frac{D+2}{2}(u-1)\bigg]\frac{d}{du} F^{(D-1)/2}_{\nu-1/2}(u).
\end{equation}
The asterisk sign in the summation of the above equation indicates that the $k=0$ component must be halved.  

For the geometry of the background under consideration, only the $\nabla_{r}\nabla_{w}$ and $\nabla_{\mu}\nabla_{\mu}$ differential operators contributes when acting on the VEV of the field squared.
The remaining contributions come from the electromagnetic covariant derivatives acting on the Wightman function. As to the azimuthal term, it is more convenient to act the $D_{\phi}D_{\phi'}^{\dagger}$ operator in \eqref{propagator-to-sum}, and subsequently take the coincidence limit in the angular variable. Following this procedure, we obtain:
\begin{equation}
	S(q,\alpha,\chi)=\sum_{n=-\infty}^{\infty}q^2(n+\alpha)^2I_{q|n+\alpha|}(\chi),
\end{equation}
where $\chi=rr'/2s^2$. This sum can be developed by using the differential equation obeyed by the modified Bessel equation. Then we get,
\begin{equation}
	S(q,\alpha,\chi)=\bigg(\chi^2\frac{d^2}{d\chi^2}+\chi\frac{d}{d\chi}-\chi^2\bigg)\sum_{n=-\infty}^{\infty}I_{q|n+\alpha|}(\chi),
\end{equation}
where this last sum is given by
\begin{equation}
	\sum_{n=-\infty}^{\infty}I_{q|n+\alpha|}(\chi)=\frac{2}{q}\sideset{}{'}\sum_{k=0}^{[q/2]}\cos(2\pi k\alpha)e^{\chi\cos(2\pi k/q)}-\frac{2}{\pi}\int_{0}^{\infty}dy\frac{e^{-\chi\cosh(2y)}f(q,\varepsilon,2y)}{\cosh(2qy)-\cos(q\pi)}.
\end{equation}

The compactified induced contribution in the VEV of the energy-momentum tensor is calculated by making use of the corresponding parts in the Wightman function and VEV of the field squared. After long but straightforward steps, we get (no summation over $\mu$)
\begin{eqnarray}
	\langle T_{\mu}^{\mu}\rangle_{\rm{c}}&=&-\frac{8}{(2\pi)^{\frac{D+1}{2}}a^{D+1}}\sum_{l=1}^{\infty}\cos(2\pi \tilde{\beta}l)\Bigg[\sideset{}{'_*}\sum_{k=0}^{[q/2]}\cos(2\pi k\varepsilon)g_{\mu}^{(l)}(v_{lk},\sin(\pi k/q))\nonumber\\
	&-&\frac{q}{\pi}\int_{0}^{\infty}dy\frac{f(q,\varepsilon,2y)g_{\mu}^{(l)}(v_{ly},\cosh(y))}{\cosh(2qy)-\cos(q\pi)}\Bigg]  \  ,
	\label{E-M-compactification}
\end{eqnarray}
where we have defined the function
\begin{equation}
	g_{\mu}^{(l)}(u,v)=G_{\mu,l}^{\mu}(u,v)+(4\xi-1)\bigg[g(u,v)-\frac{(lLv)^2}{w^2}\frac{d^2}{du^2}F^{(D-1)/2}_{\nu-1/2}(u)\bigg]-\xi DF^{(D-1)/2}_{\nu-1/2}(u) \ .
	\label{gfunction}
\end{equation}
The function $G_{\mu,l}^{\mu}$ for separate components are given by the component expressions
\begin{eqnarray}
	G_{0,l}^{0}(u,v)&=&-[1+2\xi (u-1)]\frac{d}{du}F^{(D-1)/2}_{\nu-1/2}(u)\nonumber\\
	G_{1,l}^{1}(u,v)&=&[2v^2(1-2\xi)-1-2\xi (u-1)]\frac{d}{du}F^{(D-1)/2}_{\nu-1/2}(u)\nonumber\\
	&+&2v^2(1-4\xi)\bigg[u-1-\frac{(lL)^2}{2w^2}\bigg]\frac{d^2}{du^2}F^{(D-1)/2}_{\nu-1/2}(u)\nonumber\\
	G_{2,l}^{2}(u,v)&=&-[1+2v^2(2\xi-1)+2\xi(u-1)]\frac{d}{du}F^{(D-1)/2}_{\nu-1/2}(u)\nonumber\\&-&2(1-v^2)\bigg[u-1-\frac{(lL)^2}{2w^2}\bigg]\frac{d^2}{du^2}F^{(D-1)/2}_{\nu-1/2}(u)\nonumber\\
	G_{3,l}^{3}(u,v)&=&[(1-4\xi)(u-1)-1]\frac{d}{du}F^{(D-1)/2}_{\nu-1/2}(u)+(u-1)^2(1-4\xi)\frac{d^2}{du^2}F^{(D-1)/2}_{\nu-1/2}(u)\nonumber\\
	G_{4,l}^{4}(u,v)&=&G_{0,l}^{0}(u,v)-\frac{(lL)^2}{w^2}\frac{d^2}{du^2}F^{(D-1)/2}_{\nu-1/2}(u) \  .
	\label{Gfunction}
\end{eqnarray}
For the components $\mu=5,...,D$, associated with the noncompactified extra dimensions, we have (no summation) $\langle T_{\mu}^{\mu}\rangle_{\rm{c}}=\langle T_{0}^{0}\rangle_{\rm{c}}$.
This is a consequence of the invariance of the problem with respect to the boost along the corresponding directions.

\begin{figure}[!htb]
	\begin{center}
		\includegraphics[scale=0.45]{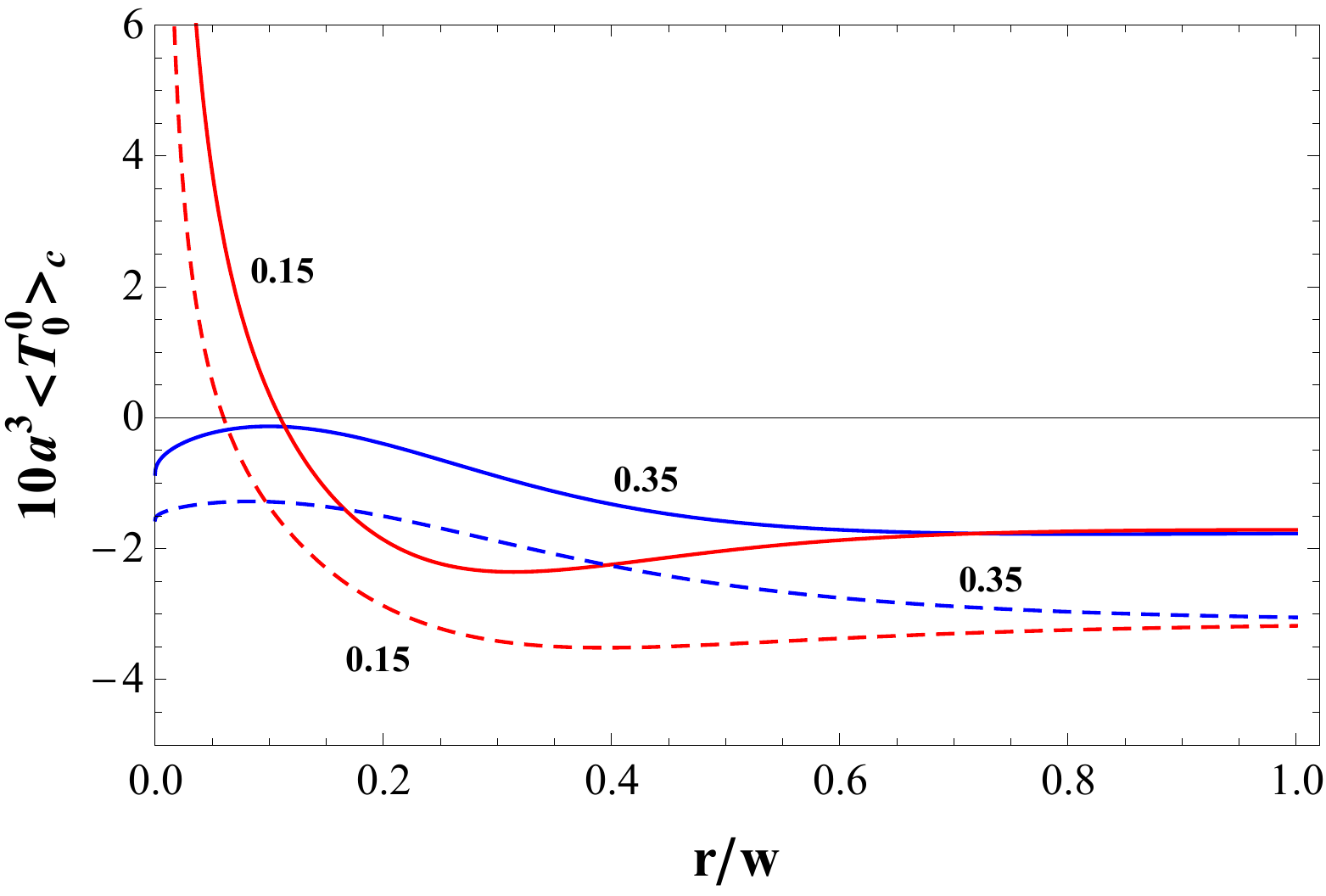}
		\quad
		\includegraphics[scale=0.45]{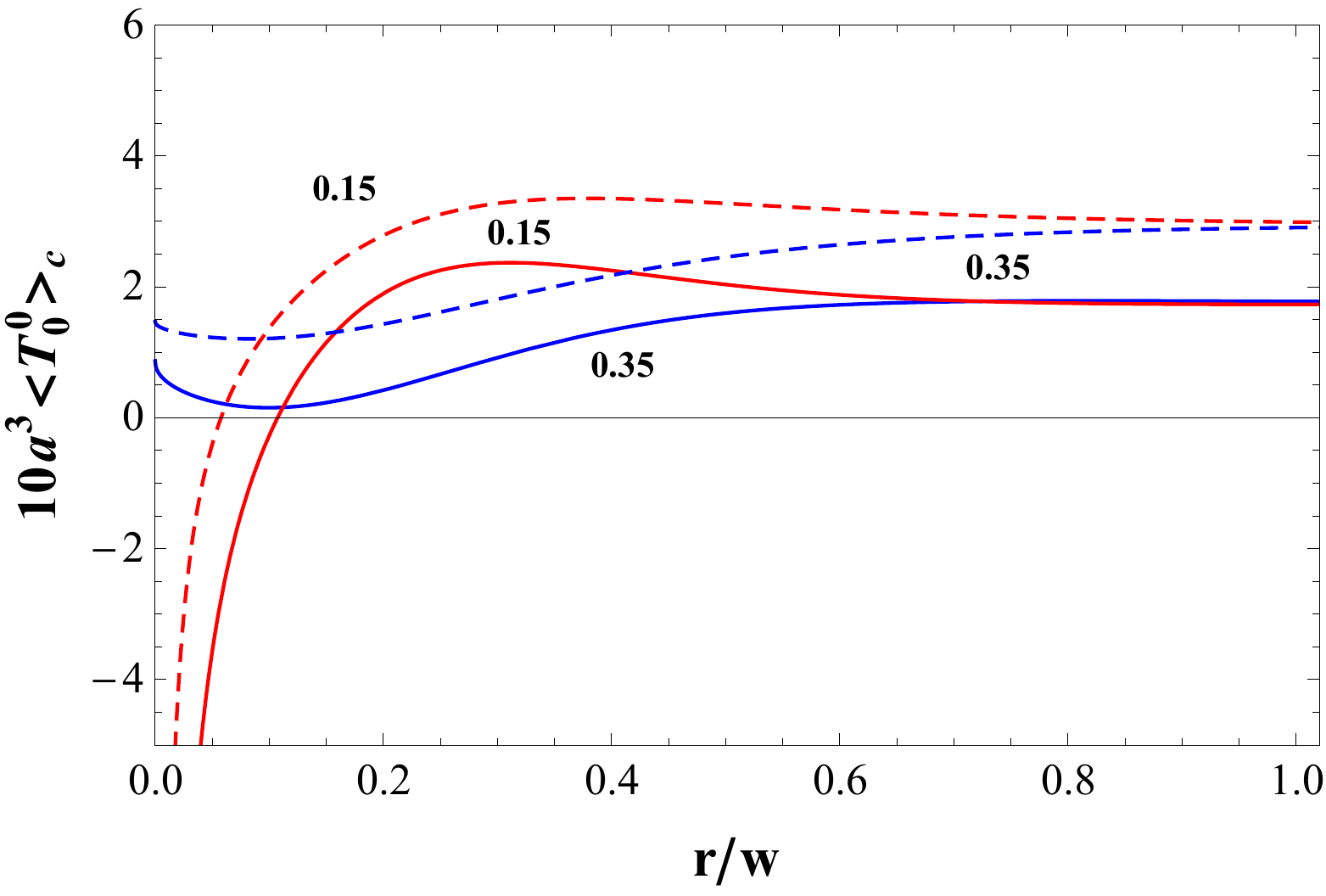}
		\caption{The VEV energy density component associated with the compactification is plotted for $D=4$, in units of $a^{3}$, as a function of $r/w$,  for $q=3.5$, $L/w=1$ and $m=0$. The numbers near the curves correspond to the values of parameter $\varepsilon$. The left panel corresponds to $\tilde{\beta}=0$ and the right one to $\tilde{\beta}=0.5$. The full lines correspond to the massless minimum coupling field while the dashed ones correspond to the conformal massless case.}
		\label{fig3a-b}
	\end{center}
\end{figure}

We want now to investigate the asymptotic behavior of the energy-density component associated with the compactification, $\langle T_{0}^{0}\rangle_{\rm{c}}$, for points near the string's core. By analyzing the structure of the terms in \eqref{gfunction} and \eqref{Gfunction}, and after some intermediate steps, we can observe that for values of $q>1/|\varepsilon|$, the energy-density is finite at $r=0$. However, for values of $q<1/|\varepsilon|$, the energy-density is divergent on the string's core. This divergence comes from the integral part of \eqref{E-M-compactification} for the corresponding component of the energy-moment tensor. The main reason to the appearance of the divergence is in the function $f(q,\varepsilon,2y)$ defined in \eqref{ffunction}. If there $\varepsilon=0$, there are no hyperbolic cosines involving the variable of integration, $y$, and the integral is always finite. In order to find how this divergence appears  we will adopt the following procedure: For large values of $y$ we take the approximation $\cosh(y)\approx e^y/2$. Analyzing the integral part of $\langle T_{0}^{0}\rangle_{\rm{c}}$, for $D=4$, we found that near the string the divergent term behaves as $(w/r)^\sigma$, with $\sigma=2(1-q|\varepsilon|)$. In Fig.\ref{fig3a-b} we have plotted the energy density component for a fixed $q$ and different values of $\varepsilon$, for the minimum and conformal coupled massless quantum scalar field. By this graph we can see that for $\varepsilon$ smaller th $1/q$ the behavior of $\langle T_{0}^{0}\rangle_{\rm{c}}$ at origin is divergent, for both cases of $\tilde{\beta}$. A similar analysis could also be developed for the other components of the energy-momentum tensor.
%

For the non-zero off-diagonal component, we have
\begin{eqnarray}
\langle T_{3}^{1}\rangle_{\rm{c}}&=&-\frac{8}{(2\pi)^{\frac{D+1}{2}}a^{D+1}}\frac{w}{r}\sum_{l=1}^{\infty}\cos(2\pi\tilde{\beta}l)\Bigg[\sideset{}{'_*}\sum_{k=0}^{[q/2]}\cos(2\pi k\varepsilon)h^{(l)}(v_{lk})\nonumber\\&-&\frac{q}{\pi}\int_{0}^{\infty}dy\frac{f(q,\varepsilon,2y)h^{(l)}(v_{ly})}{\cosh(2qy)-\cos(q\pi)}\Bigg],
\label{E-M-compact-off-diagonal}
\end{eqnarray}
where we have defined the function
\begin{equation}
h^{(l)}(u)=\bigg(u-1-\frac{(lL)^2}{2w^2}\bigg)\bigg[(1-2\xi)\frac{d}{du}F^{(D-1)/2}_{\nu-1/2}(u)+(u-1)(1-4\xi)\frac{d^2}{du^2}F^{(D-1)/2}_{\nu-1/2}(u)\bigg].
\end{equation}

It is possible to check that the compactified energy-momentum tensor, $\langle T_{\mu}^{\mu}\rangle_{\rm{c}}$, satisfies the following identity:
\begin{equation}
	\langle T_{\mu}^{\mu}\rangle_{\rm{c}}=2[D(\xi-\xi_{c})\nabla_{\mu}\nabla^{\mu}\langle|\varphi|^2\rangle_{\rm{c}}+m^2\langle|\varphi|^2\rangle_{\rm{c}}].
	\label{EM-trace}
\end{equation}
Note it is traceless for a conformal massless quantum scalar field.
Furthermore, as an additional verification for the expressions in \eqref{E-M-compactification} and off-diagonal component in \eqref{E-M-compact-off-diagonal}, we can see that the covariant conservation equation $ \nabla_{\mu}\langle T_{\nu}^{\mu}\rangle=0$ is obeyed. For the geometry under consideration the latter is reduced to the following relations
\begin{equation}
	\frac{1}{r}\partial_{r}(r\langle T_{1}^{1}\rangle_{\rm{c}})-\frac{1}{r}\langle T_{2}^{2}\rangle_{\rm{c}}-\frac{D+1}{w}\langle T_{1}^{3}\rangle_{\rm{c}}+\partial_{w}\langle T_{1}^{3}\rangle_{\rm{c}}=0
	\label{c-eq-1}
\end{equation}
and
\begin{equation}
	\frac{1}{r}\partial_{r}(r\langle T_{3}^{1}\rangle_{\rm{c}})+\partial_{w}\langle T_{3}^{3}\rangle_{\rm{c}}-\frac{D+1}{w}\langle T_{3}^{3}\rangle_{\rm{c}}+\frac{1}{w}\langle T_{\mu}^{\mu}\rangle_{\rm{c}}=0.
	\label{c-eq-2}
\end{equation}

For a conformally coupled massless scalar field, the energy density component, $\langle T_{0}^{0}\rangle_{\rm{c}}$, is
\begin{eqnarray}
\langle T_{0}^{0}\rangle_{\rm{c}}&=&\frac{4\Gamma(\frac{D+1}{2})}{(4\pi)^{\frac{D+1}{2}}Da^{D+1}}\sum_{l=1}^{\infty}\cos(2\pi \tilde{\beta}l)\Bigg\{\sideset{}{'_*}\sum_{k=0}^{[q/2]}\cos(2\pi k\varepsilon)\Bigg[Df^{\frac{{D-1}}{2}}(\rho_{lk})\nonumber\\&-&\Bigg(D+2s_{k}^2+(2D+1)\rho_{lk}\Bigg)f^{\frac{D+1}{2}}(\rho_{lk})+(D+1)\Bigg(\frac{r^2s_{k}^4}{w^2}+\rho_{lk}^2\Bigg)f^{\frac{D+3}{2}}(\rho_{lk})\Bigg]\nonumber\\&-&\frac{q}{\pi}\int_{0}^{\infty}dy\frac{f(q,\varepsilon,2y)}{\cosh(2qy)-\cos(q\pi)}\Bigg[Df^{\frac{{D-1}}{2}}(\rho_{ly})-\Bigg(D+2s_{y}^2+(2D+1)\rho_{ly}\Bigg)\nonumber\\&\times&f^{\frac{D+1}{2}}(\rho_{ly})+(D+1)\Bigg(\frac{r^2s_{y}^4}{w^2}+\rho_{ly}^2\Bigg)f^{\frac{D+3}{2}}(\rho_{ly})\Bigg]\Bigg\} \ ,
\label{EMT-comp00-conformal}
\end{eqnarray}
where we have introduced the function
\begin{equation}
f^{\alpha}(\rho)=\frac{1}{\rho^{\alpha}}-\frac{1}{(1+\rho)^{\alpha}}
\label{ffunction2}
\end{equation}
and the variable
\begin{eqnarray}
\rho_{l\gamma}&=&\bigg(\frac{rs_{\gamma}}{w}\bigg)^2+\bigg(\frac{lL}{2w}\bigg)^2 \ ,
\label{rho-var}
\end{eqnarray}
with $s_{k}=\sin(k\pi/q)$ and $s_{y}=\cosh(y)$.
 
The off-diagonal component for the conformal coupled massless scalar quantum field is given by
\begin{eqnarray}
\langle T_{3}^{1}\rangle_{\rm{c}}&=&-\frac{8\Gamma(\frac{D+3}{2})}{(4\pi)^{\frac{D+1}{2}}Da^{D+1}}\frac{r}{w}\sum_{l=1}^{\infty}\cos(2\pi\tilde{\beta}l)\Bigg\{\sideset{}{'_*}\sum_{k=0}^{[q/2]}\cos(2\pi k\varepsilon)\sin^2(\pi k/q)\nonumber\\&\times&\bigg[1+\frac{r^2}{w^2}\sin^{2}(\pi k/q)+\frac{(lL)^2}{4w^2}\bigg]^{-\frac{D+3}{2}}-\frac{q}{\pi}\int_{0}^{\infty}dy\frac{f(q,\varepsilon,2y)\cosh^2(y)}{\cosh(2qy)-\cos(q\pi)}\nonumber\\&\times&\bigg[1+\frac{r^2}{w^2}\cosh^{2}(y)+\frac{(lL)^2}{4w^2}\bigg]^{-\frac{D+3}{2}}\Bigg\} \ .
\label{conformal-off-diagonal-c}
\end{eqnarray}

The cosmic string part in the VEV of the energy-momentum tensor is calculated from \eqref{E-M-Tensor-formula}, by making use of the string component of the Wightman function and VEV of the field squared given in \eqref{phi-squared}. After long but straightforward calculations, for the string part, one finds (no summation over $\mu$)
\begin{eqnarray}
	\langle T_{\mu}^{\mu}\rangle_{\rm{cs}}&=&-\frac{4}{(2\pi)^{\frac{D+1}{2}}a^{D+1}}\Bigg[\sideset{}{'}\sum_{k=1}^{[q/2]}\cos(2\pi k\varepsilon)g_{\mu}^{(0)}(v_{0k},\sin(\pi k/q))\nonumber\\
	&-&\frac{q}{\pi}\int_{0}^{\infty}dy\frac{f(q,\varepsilon,2y)g_{\mu}^{(0)}(v_{0y},\cosh(y))}{\cosh(2qy)-\cos(q\pi)}\Bigg],
	\label{E-M-cosmic-string}
\end{eqnarray}
where
\begin{equation}
	g_{\mu}^{(0)}(u,v)=G_{\mu,0}^{\mu}(u,v)+(4\xi-1)g(u,v)-\xi DF^{(D-1)/2}_{\nu-1/2}(u),
\end{equation}
and
\begin{eqnarray}
	G_{0,0}^{0}(u,v)&=&-[1+2\xi (u-1)]\frac{d}{du}F^{(D-1)/2}_{\nu-1/2}(u)\nonumber\\
	G_{1,0}^{1}(u,v)&=&[2v^2(1-2\xi)-1-2\xi (u-1)]\frac{d}{du}F^{(D-1)/2}_{\nu-1/2}(u)\nonumber\\&+&2v^2(1-4\xi)(u-1)\frac{d^2}{du^2}F^{(D-1)/2}_{\nu-1/2}(u)\nonumber\\
	G_{2,0}^{2}(u,v)&=&-[1+2v^2(2\xi-1)+2\xi(u-1)]\frac{d}{du}F^{(D-1)/2}_{\nu-1/2}(u)\nonumber\\&-&2(u-1)(1-v^2)\frac{d^2}{du^2}F^{(D-1)/2}_{\nu-1/2}(u)\nonumber\\
	G_{3,0}^{3}(u,v)&=&[(1-4\xi)(u-1)-1]\frac{d}{du}F^{(D-1)/2}_{\nu-1/2}(u)\nonumber\\&+&(u-1)^2(1-4\xi)\frac{d^2}{du^2}F^{(D-1)/2}_{\nu-1/2}(u).
\end{eqnarray}
As to the other components with $\mu=4,...,D$, we have (no summation) $\langle T_{\mu}^{\mu}\rangle_{\rm{cs}}=\langle T_{0}^{0}\rangle_{\rm{cs}}$. 
  
For the non-zero off-diagonal component, we have
\begin{eqnarray}
	\langle T_{3}^{1}\rangle_{\rm{cs}}=-\frac{4}{(2\pi)^{\frac{D+1}{2}}a^{D+1}}\frac{w}{r}\Bigg[\sideset{}{'}\sum_{k=1}^{[q/2]}\cos(2\pi k\varepsilon)h^{(0)}(v_{k})-\frac{q}{\pi}\int_{0}^{\infty}dy\frac{f(q,\varepsilon,2y)h^{(0)}(v_{y})}{\cosh(2qy)-\cos(q\pi)}\Bigg],
	\label{E-M-cs-off-diagonal}
\end{eqnarray}
where
\begin{equation}
	h^{(0)}(u)=(u-1)\bigg[(1-2\xi)\frac{d}{du}F^{(D-1)/2}_{\nu-1/2}(u)+(u-1)(1-4\xi)\frac{d^2}{du^2}F^{(D-1)/2}_{\nu-1/2}(u)\bigg].
\end{equation}
It is possible to check that the expressions in \eqref{E-M-cosmic-string} obey the traceless condition for a conformal coupled massless field. The expressions given in \eqref{E-M-cosmic-string}and \eqref{E-M-cs-off-diagonal} also satisfy the continuity equations as those ones given in \eqref{c-eq-1} and \eqref{c-eq-2}.

For a conformally coupled massless scalar field, the energy density component, $\langle T_{0}^{0}\rangle_{\rm{cs}}$, is given by
\begin{eqnarray}
\langle T_{0}^{0}\rangle_{\rm{cs}}&=&\frac{2\Gamma(\frac{D+1}{2})}{(4\pi)^{\frac{D+1}{2}}Da^{D+1}}\Bigg\{\sideset{}{'}\sum_{k=1}^{[q/2]}\cos(2\pi k\varepsilon)\Bigg[Df^{\frac{{D-1}}{2}}(\rho_{0k})\nonumber\\&-&\Bigg(D+2s_{k}^2+(2D+1)\rho_{0k}\Bigg)f^{\frac{D+1}{2}}(\rho_{0k})+(D+1)\Bigg(\frac{r^2s_{k}^4}{w^2}+\rho_{0k}^2\Bigg)f^{\frac{D+3}{2}}(\rho_{0k})\Bigg]\nonumber\\&-&\frac{q}{\pi}\int_{0}^{\infty}dy\frac{f(q,\varepsilon,2y)}{\cosh(2qy)-\cos(q\pi)}\Bigg[Df^{\frac{{D-1}}{2}}(\rho_{0y})-\Bigg(D+2s_{y}^2+(2D+1)\rho_{0y}\Bigg)\nonumber\\&\times&f^{\frac{D+1}{2}}(\rho_{0y})+(D+1)\Bigg(\frac{r^2s_{y}^4}{w^2}+\rho_{0y}^2\Bigg)f^{\frac{D+3}{2}}(\rho_{0y})\Bigg]\Bigg\} \ ,
\label{EMT-cs-comp00-conformal}
\end{eqnarray}
where $\rho_{0k}$ and $\rho_{0y}$ are given in \eqref{rho-var} for $l=0$, and the function $f^{\alpha}(\rho)$ is defined in \eqref{ffunction2}.
\begin{figure}[!htb]
	\begin{center}
		\centering
		\includegraphics[scale=0.5]{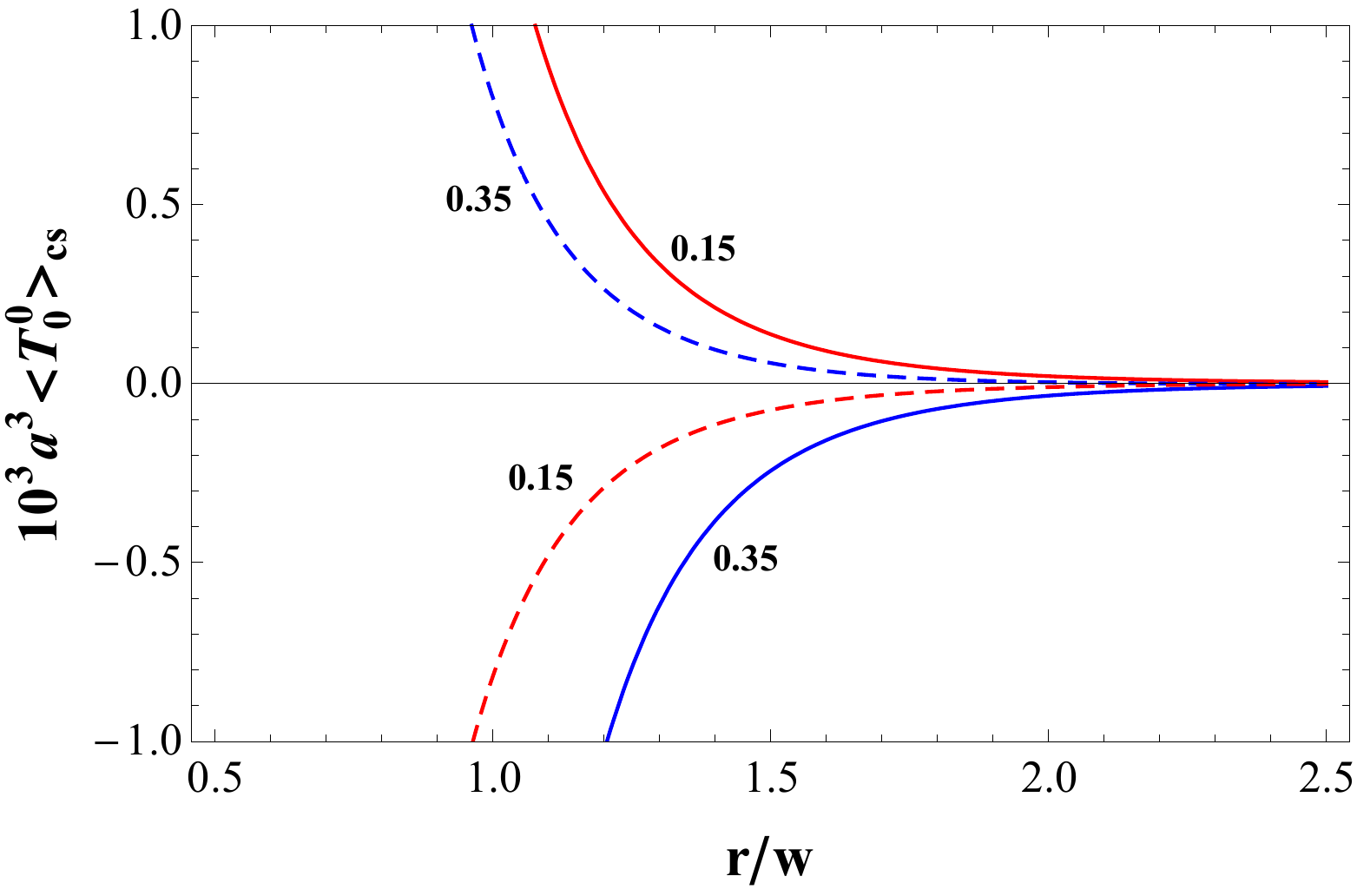}
		\caption{The VEV of the energy density in uncompactified string geometry is plotted for $D=4$, in units of $a^{3}$, in terms of $r/w$,  for $q=2.5$ and $m=0$. The numbers near the curves refer to the values of the parameter $\varepsilon$. The full lines correspond to the massless minimum coupling field while the dashed ones correspond to the conformal massless case.}
		\label{fig4}
	\end{center}
\end{figure}

In Fig.\ref{fig4} we have plotted the energy density in uncompactified cosmic string geometry, $\langle T_{0}^{0}\rangle_{\rm{cs}}$, as function of  the proper distance from the string measured in units of the AdS curvature radius $a$. We note from this figure that the parameters associated with magnetic flux along the string, $\varepsilon$, and the curvature coupling, $\xi$, can effectively change the intensity and behavior of the energy density.

For the conformal massless case, the component off-diagonal of the energy-momentum tensor, $\langle T_{3}^{1}\rangle_{\rm{cs}}$, is written as
\begin{eqnarray}
\langle T_{3}^{1}\rangle_{\rm{cs}}&=&-\frac{4\Gamma(\frac{D+3}{2})}{(4\pi)^{\frac{D+1}{2}}Da^{D+1}}\frac{r}{w}\Bigg\{\sideset{}{'}\sum_{k=1}^{[q/2]}\cos(2\pi k\varepsilon)\sin^2(\pi k/q)\nonumber\\&\times&\bigg[1+\frac{r^2}{w^2}\sin^{2}(\pi k/q)\bigg]^{-\frac{D+3}{2}}-\frac{q}{\pi}\int_{0}^{\infty}dy\frac{f(q,\varepsilon,2y)\cosh^2(y)}{\cosh(2qy)-\cos(q\pi)}\nonumber\\&\times&\bigg[1+\frac{r^2}{w^2}\cosh^{2}(y)\bigg]^{-\frac{D+3}{2}}\Bigg\} \ .
\label{conformal-off-diagonal-cs}
\end{eqnarray}
As you can see in the conformal coupled massless scalar field, the above equation is expressed in terms of elementary functions. 
\section{Concluding remarks}
\label{concl}

In this paper, we have investigated the vacuum polarization associated to a massive scalar charged quantum field in a high dimension AdS spacetime in the presence of a  cosmic string. In this analysis we also admit that an extra coordinate is compactified to a circle, and two distinct magnetic fluxes are present. One running along the string's core and the other is along the center of the compactified dimension. In this way four different ingredients take place to produce contributions in the vacuum polarization phenomenon. Specifically we have computed explicitly the VEV of the field squared and the energy-momentum tensor. In order to develop these calculations, we have constructed the positive Wightman function, by using the mode sum of positive frequency solution of the Klein-Gordon equation, Eq. \eqref{COS}. This function whose closed form is presented in Eq. \eqref{full-propagator}, has been shown to be decomposed in three contributionsas \eqref{wightman-function-expanded}. The first one due to the pure AdS bulk, the second associated to the cosmic string in AdS space, and the third one, associated with the compactification.

In our analysis of the  VEV of field squared, we have concentrated in the contributions due to the cosmic string and the compactification. As to the string induced part, it is  presented in closed form in \eqref{phi-squared}. We have investigated this result for small and great values of the ratio $r/w$. These asymptotic behaviors are given in \eqref{field-squared-cs-assymp} and \eqref{field-squared-cs-assymp2}, respectively. The expression for the conformal coupled massless scalar field is presented in \eqref{conformal-phi-squared}. In Fig.\ref{fig1} the string induced part is plotted for $D=4$, in units of $a^3$, as a function of the proper distance form the string in units of AdS curvature radius, $r/w$. In this plot is showed that the magnetic flux running along the string's core can modify the behavior and the magnitude of $\langle |\phi|^2\rangle_{\rm{cs}}$. As to the VEV of the field squared associated with the compact extra dimension, the closed expression is presented in \eqref{field-squared-c}. This contribution presents two parts: The first one, Eq. \eqref{term-k-0}, is the term $k=0$ which depends only on parameters associated with the compactification. As to the second set, \eqref{VEV-field-squared-c}, it depends on the planar angular deficit induced by the cosmic string, $q,$ and on the magnetic flux running through its core, $\varepsilon$. The asymptotic expression for great values of $L/w$ is given in \eqref{assymp-field-squared-c}. The conformal coupled massless scalar field case is presented in \eqref{field-squared-c-conformal}. In Fig.\ref{fig2} we have plotted Eq.\eqref{field-squared-c} for $D=4$, in units of $a^3$, as function of $L/w$. From this plot we note that the parameter $\tilde{\beta}$ can increase or decrease the magnitude of the VEV contribution induced by the compactification, as well to change its behavior. We also note that this VEV decrease as the length of compact extra dimension, $L$, increase.

Another important analysis developed in this paper is the VEV of the energy-momentum tensor. This observable also presents three distinct contributions as exhibited in \eqref{TEM0}. In this paper we were mainly interested in the contribution due to the cosmic string and the compactification, given by Eq.s \eqref{E-M-cosmic-string} and \eqref{E-M-compactification}, respectively. We have checked that both contributions satisfy the covariant conservation equation and their trace are related to the respective VEVs of the field squared by \eqref{EM-trace}. As to the compactified induced contribution, the energy density component, $\langle T_{0}^{0}\rangle_{\rm{c}}$, is plotted in Fig.\ref{fig3a-b} for $D=4$, as function of $r/w$ with different values of $\varepsilon$. From this figure we note that it is finite for $q>1/|\varepsilon|$ and it is divergent for values of $q<1/|\varepsilon|$, on the string's core, $r=0$, as we have previously observed analytically. We also can note from this plot the influence of the parameter $\tilde{\beta}$ on the behavior of the energy density. The off-diagonal component of the compact extra dimension part is given in \eqref{E-M-compact-off-diagonal} and its expression for the conformal coupled massless scalar field is given in \eqref{conformal-off-diagonal-c}. As to the cosmic string induced part, the energy density for the conformal coupled massless scalar field is presented in \eqref{EMT-cs-comp00-conformal}. The off-diagonal component is given in \eqref{E-M-cs-off-diagonal} and its correspondent conformal coupled massless scalar field expression is presented in \eqref{conformal-off-diagonal-cs}. In Fig.\ref{fig4} is displayed the behavior of the energy density induced by the string for $D=4$, as a function of the $r/w$. From this figure we note the influence on the intensity and behavior of $\langle T_{0}^{0}\rangle_{\rm{cs}}$ by the curvature coupling and the magnetic flux running along the string's core.

Finally we would like to mention that our results for the VEVs of the field squared and  energy-momentum tensor are as general as possible. They can be applied to several particular situations. For example, if we would like to analyze only the influence of the presence of a cosmic string in a four dimension AdS spacetime, we have only to take the limit $L\to\infty$ and discard the extra coordinates. Moreover, in this paper the expression for the calculation of the vacuum expectation value of the energy-momentum tensor associated with a charged scalar quantum field in curved spacetime considering  electromagnetic interaction, Eq. \eqref{E-M-Tensor-formula}, to the best of our knowledge, is presented for the first time in literature. Besides our expression presents a very compact format. 
\section*{Acknowledgments}
W.O.S thanks CAPES for financial support. H.F.M thanks CNPq for partial financial support
under Grants no 305379/2017-8. E.R.B.M is partially supported by CNPq under Grant no
313137/2014-5.
\appendix
\section{Appendix: Energy-Momentum Tensor Formula}\label{Apendx}
In this Appendix we shall present an alternative expression to calculate the VEV of the energy-momentum tensor. 

The action associated with a charged scalar field coupled with a gauge field is given by
\begin{equation}
S=\int d^4x\sqrt{-g(x)}\big[g^{\mu\nu}(x)(D_{\mu}\varphi(x))(D_{\nu}\varphi(x))^\dagger-(m^2+\xi R)\varphi^\dagger(x)\varphi(x)\big].
\label{action}
\end{equation}
By using the definition of the energy-momentum tensor below\cite{BD},
\begin{equation}
T_{\alpha\beta}(x)=\frac{2}{\sqrt{-g}}\frac{\delta S}{\delta g^{\alpha\beta}(x)} \ ,
\label{EMT}
\end{equation}
we obtain:
\begin{eqnarray}
T_{\alpha\beta}&=&-g_{\alpha\beta}[g^{\mu\nu}(D_{\mu}\varphi)(D_{\nu}\varphi)^\dagger-(m^2+\xi R)\varphi^\dagger\varphi]+(D_{\alpha}\varphi)(D_{\beta}\varphi)^\dagger+(D_{\alpha}\varphi)^\dagger(D_{\beta}\varphi)\nonumber\\
&-&2\xi(R_{\alpha\beta}-g_{\alpha\beta}\Box+\nabla_{\alpha}\nabla_{\beta})\varphi^{\dagger}\varphi \ .
\label{ap.0a}
\end{eqnarray}
Let us develop the first term in the square bracket on the  right-hand side of the above expression:
\begin{eqnarray}
(i)=g^{\mu\nu}(\nabla_{\mu}\varphi)(\nabla_{\nu}\varphi^{\dagger})-ieA^{\mu}(\varphi^\dagger\nabla_{\mu}\varphi-\varphi\nabla_{\mu}\varphi^\dagger)+e^2A_{\mu}A^{\mu}\varphi^\dagger\varphi \ ,
\label{ap.1}
\end{eqnarray}
where we have made use of  $D_{\mu}=\nabla_{\mu}+ieA_{\mu}$. However, the operator $D_{\mu}D^{\mu}$ acting on $\varphi$ provides the expression below
\begin{equation}
D_{\mu}D^{\mu}\varphi=\Box\varphi+ie(\nabla_{\mu}A^{\mu})\varphi+2ieA^{\mu}\nabla_{\mu}\varphi-e^2A_{\mu}A^{\mu}\varphi.
\label{ap.2}
\end{equation}
Multiplying the above expression by $\varphi^\dagger$/2, and then summing it with its complex conjugate, we get
\begin{eqnarray}
\frac{1}{2}[\varphi^\dagger D_{\mu}D^{\mu}\varphi+\varphi(D_{\mu}D^{\mu}\varphi)^\dagger]&=&\frac{1}{2}(\varphi^\dagger\Box\varphi+\varphi\Box\varphi^\dagger)+ieA^{\mu}(\varphi^\dagger \nabla_{\mu}\varphi-\varphi\nabla_{\mu}\varphi^\dagger)\nonumber\\
&-&e^2A_{\mu}A^{\mu}\varphi^\dagger\varphi \ .
\label{ap.5}
\end{eqnarray}
On the other hand, we have
\begin{eqnarray}
\Box(\varphi^\dagger\varphi)=\varphi\Box\varphi^\dagger+\varphi^\dagger\Box\varphi+2g^{\mu\nu}(\nabla_{\mu}\varphi^\dagger)(\nabla_{\nu}\varphi) \ .
\label{ap.6}
\end{eqnarray}
Now, conveniently combining \eqref{ap.5} and \eqref{ap.6}, we can rewrite \eqref{ap.1} as
\begin{equation}
(i)=\frac{1}{2}\Box(\varphi^\dagger\varphi)-\frac{1}{2}[\varphi^\dagger D_{\mu}D^{\mu}\varphi+\varphi(D_{\mu}D^{\mu}\varphi)^\dagger] \ .
\label{ap.7}
\end{equation}
Multiplying  $\varphi^{\dagger}$ on the left-hand side of the equation of motion for $\varphi$, we obtain
\begin{equation}
\varphi^\dagger D_{\mu}D^{\mu}\varphi=-(m^2+\xi R)\varphi^\dagger\varphi \ .
\label{ap.8}
\end{equation}
Summing the above equation with its complex conjugate, apart of a factor $1/2$, we have
\begin{equation}
\frac{1}{2}[\varphi^\dagger D_{\mu}D^{\mu}\varphi+\varphi(D_{\mu}D^{\mu}\varphi)^\dagger]=-(m^2+\xi R)\varphi^\dagger\varphi \ .
\label{ap.9}
\end{equation}
Substituting \eqref{ap.9} into \eqref{ap.7}, we get
\begin{equation}
(i)=\frac{1}{2}\Box(\varphi^\dagger\varphi)+(m^2+\xi R)\varphi^\dagger\varphi \ .
\label{ap.10}
\end{equation}
Taking the above expression into \eqref{ap.0a}, the energy-momentum tensor can be written as,
\begin{eqnarray}
T_{\alpha\beta}=(D_{\alpha}\varphi)(D_{\beta}\varphi)^\dagger+(D_{\alpha}\varphi)^\dagger(D_{\beta}\varphi)-2[\xi R_{\alpha\beta}+\xi\nabla_{\alpha}\nabla_{\beta}-(\xi-1/4)g_{\alpha\beta}\Box]\varphi^{\dagger}\varphi \ .
\end{eqnarray}
Hence, the VEV of the operator stress-energy tensor found is formally written as
\begin{equation}
\langle T_{\alpha\beta}\rangle=(D_{\alpha}D_{\beta'}^{\dagger}+D_{\alpha'}^{\dagger}D_{\beta})W(x,x')-2[\xi R_{\alpha\beta}+\xi\nabla_{\alpha}\nabla_{\beta}-(\xi-1/4)g_{\alpha\beta}\nabla_{\sigma}\nabla^{\sigma}]\langle|\varphi|^2\rangle.
\label{E-M-Tensor-formula-2}
\end{equation}
	
\end{document}